\documentclass[aps,prl,twocolumn,superscriptaddress]{revtex4-2}
\usepackage[english]{babel}
\usepackage{geometry,amsmath,amssymb,graphicx,textcomp,xcolor,natbib}
\usepackage[utf8]{inputenc}
\usepackage{multirow}
\newcommand{\mathd}{\mathrm{d}}
\newcommand{\mathe}{\mathrm{e}}

\newcommand{\qlk}{{\sc QuaLiKiz}}
\newcommand{\qlkjet}{{\sc QuaLiKiz--Jetto}}
\newcommand{\gys}{{\sc Gysela}}
\newcommand{\gkw}{{\sc Gkw}}

\newcommand{\exb}{\mathbf{E}\times \mathbf{B}}
\renewcommand{\section}[1]{\noindent{\bf #1 --}~}

\newcommand{\newstuff}[1]{\color{black}{#1}\color{black}}

{
\makeatletter
\def\frontmatter@thefootnote{%
 \altaffilletter@sw{\@fnsymbol}{\@fnsymbol}{\csname c@\@mpfn\endcsname}%
}%
\makeatother

\begin{document}
\title{
{\color{black}The Problem of Marginality in Model Reductions of Turbulence 
} }

\author{C. Gillot}
\affiliation{CEA, IRFM, F-13108 Saint-Paul-lez-Durance, France}
\author{G. Dif-Pradalier}
\email{guilhem.dif-pradalier@cea.fr}
\affiliation{CEA, IRFM, F-13108 Saint-Paul-lez-Durance, France}
\author{Y. Sarazin}
\affiliation{CEA, IRFM, F-13108 Saint-Paul-lez-Durance, France}
\author{C. Bourdelle}
\affiliation{CEA, IRFM, F-13108 Saint-Paul-lez-Durance, France}
\author{Y. Camenen}
\affiliation{Aix-Marseille Universit{\'e}, CNRS PIIM, UMR 7345 Marseille, 	France}
\author{J. Citrin}
\affiliation{DIFFER—Dutch Institute for Fundamental Energy Research, Eindhoven, Netherlands}
\author{X. Garbet}
\affiliation{CEA, IRFM, F-13108 Saint-Paul-lez-Durance, France}
\author{Ph. Ghendrih}
\affiliation{CEA, IRFM, F-13108 Saint-Paul-lez-Durance, France}
\author{V. Grandgirard}
\affiliation{CEA, IRFM, F-13108 Saint-Paul-lez-Durance, France}
\author{P. Manas}
\affiliation{CEA, IRFM, F-13108 Saint-Paul-lez-Durance, France}
\author{F. Widmer}
\affiliation{Aix-Marseille Universit{\'e}, CNRS PIIM, UMR 7345 Marseille, 	France}

\date{\today}
\begin{abstract}
Reduced quasilinear (QL) and nonlinear (gradient-driven) models with scale separations, commonly used to interpret experiments and to forecast turbulent transport levels in magnetised plasmas are tested against nonlinear models without scale separations (flux-driven). Two distinct regimes of turbulence --either far above threshold or near marginal stability-- are investigated with Boltzmann electrons. The success of reduced models especially hinges on the reproduction of nonlinear fluxes. Good agreement between models is found above threshold whilst reduced models would significantly underpredict fluxes near marginality, overlooking mesoscale flow organisation and turbulence self-advection. Constructive prescriptions whereby to improve reduced models is discussed.
\end{abstract}

{\maketitle}

\section{Introduction}
Fusion plasmas display the property, common in dynamical systems that upon surpassing a critical threshold, an instability may promptly build up, inducing large fluxes which deplete the driving gradients and inhibit the instability. Background gradients thus hover in the vicinity of nonlinear {\it near marginal} thresholds \cite{Diamond_1995}. 
Many strategies have been devised in modelling to mimic natural processes. All are not equivalent and different choices may critically affect the nature of computed statistical equilibria. 

Forcing-dependent steady-states have indeed long been observed in a variety of systems. Systems with long-range interactions, either controlled in energy or in temperature display different equilibria \cite{Thirring_1970, Ellis_2000}. Swirling flows controlled through either imposed torque or imposed velocity display distinct steady states, as well as different dynamical regimes \cite{Saint-Michel_2013}. In magnetised fusion plasmas, auxiliary heating and current drive force the system out of equilibrium through a constant flux. Mimicking nature, "flux-driven" (FD) forcing adiabatically imparts a volumetric flux to the system whose gradients self-consistently adapt, leading to the observation of large-scale transport events such as avalanches \cite{Newman_1996, Garbet_1998, Beyer_2000, McMillan_2009, Idomura_2009, Sarazin_2010, Singh_PoP20}, nonlinear structures such as zonal mean flows \cite{Kim_2003, Diamond_2005} or secondary patterns such as staircases \cite{Dif-Pradalier_2010, Dif-Pradalier_2015, Rath_2016, Peeters_2016, Dif-Pradalier_2017, Ashourvan_2019}. Such observations are absent or impaired when the system is driven through a body force, which amounts to imposing fixed mean gradients. This "gradient-driven" (GD) strategy is widely used in direct numerical computations for it is computationally efficient. It indeed enforces, with respect to the FD approach, spatial and temporal scale separations between equilibrium and fluctuations and solves for the fluctuations only.

Known differences between FD and GD frameworks have been documented \cite{Nakata_2013, Peeters_2016, Dif-Pradalier_2017}. Whether these are of practical incidence in fusion-relevant configurations is non trivial. The matter is important for there are increasing requirements for fast, reduced, yet reliable models to explore the vast parameter space of magnetised plasma turbulence, interpret experimental results and forecast future large experiments such as ITER. Currently, the more advanced reduced models are based on quasilinear theory (QLT) \cite{Laval_2018,Vedenov_1962, Drummond_1962}. With the advent of machine learning techniques, the ubiquitous closure problem of QLT is approached through data-driven techniques that use large-scale databases of first-principles GD computations. Systematic shortcomings, if any, within reference GD strategies are thus likely to be carried over to the reduced models. Comprehensive understanding of discrepancies between first-principles FD, GD and quasilinear approaches is thus important and timely, and the topic of this Letter. 

To this end, we confront reference results from nonlinear FD gyrokinetics using the \gys\ framework \cite{Grandgirard_2016} to state-of-the-art GD nonlinear gyrokinetics and GD quasilinear calculations, using respectively \gkw\ \cite{Peeters_2009} and \qlk\ \cite{Bourdelle_2007, Citrin_2017}. We further complement the study with confrontation with the QL flux-driven \qlkjet\ framework. Several instances of \qlk\ locally compute at various radii flux-surface-averaged transport coefficients which are passed on to the \textsc{Jetto} integrated modelling suite \cite{Romanelli_14} and used to evolve profiles through flux-driven transport equations. After a transport timescale, new profiles are fed to \qlk\ and the process loops. 

Main results are: (i) steady-state predictions of fluxes moderately depend on the nature of forcing {\it well above nonlinear threshold}; (ii) {\it near marginality} however, nonlinear and quasilinear GD models sizeably underpredict turbulent heat transport. Under a driving flux, profiles display 'stiffness', i.e. hover in the vicinity of their near-threshold flux-matching values. Large, hot devices such as ITER are expected to be stiff due to the temperature dependence of the gyroBohm heat flux scaling, making near marginality a regime which models must confront. Proximity to nonlinear thresholds implies additional complexity as it favours secondary pattern formation and mesoscale organisation. Despite this additional complexity, (iii) the underlying assumptions of QLT hold well across nonlinear regimes. We show that transport underprediction rather stems from the {\it choice of closure}, i.e. the {\it nonlinear saturation rule}. This work stresses the relevance of QLT for model reductions of turbulence whilst providing guidelines whereby reduced models can be improved. The nonlinear and QL approaches tested here are {the} current workhorse for estimating transport and confinement in turbulent fusion plasmas. This work thus has implications for present-day experimental data analysis and scenario extrapolation for fusion production. Novel saturation rules should strive to incorporate near marginal flux-driven specificities, often dubbed turbulence spreading \cite{Hahm_PoP05, Singh_PoP20, DifPradalier_encours22}, transport nonlocality or staircase organisation.

\section{Two distinct regimes}
\begin{figure}[t]
	\resizebox{\linewidth}{!}{\includegraphics{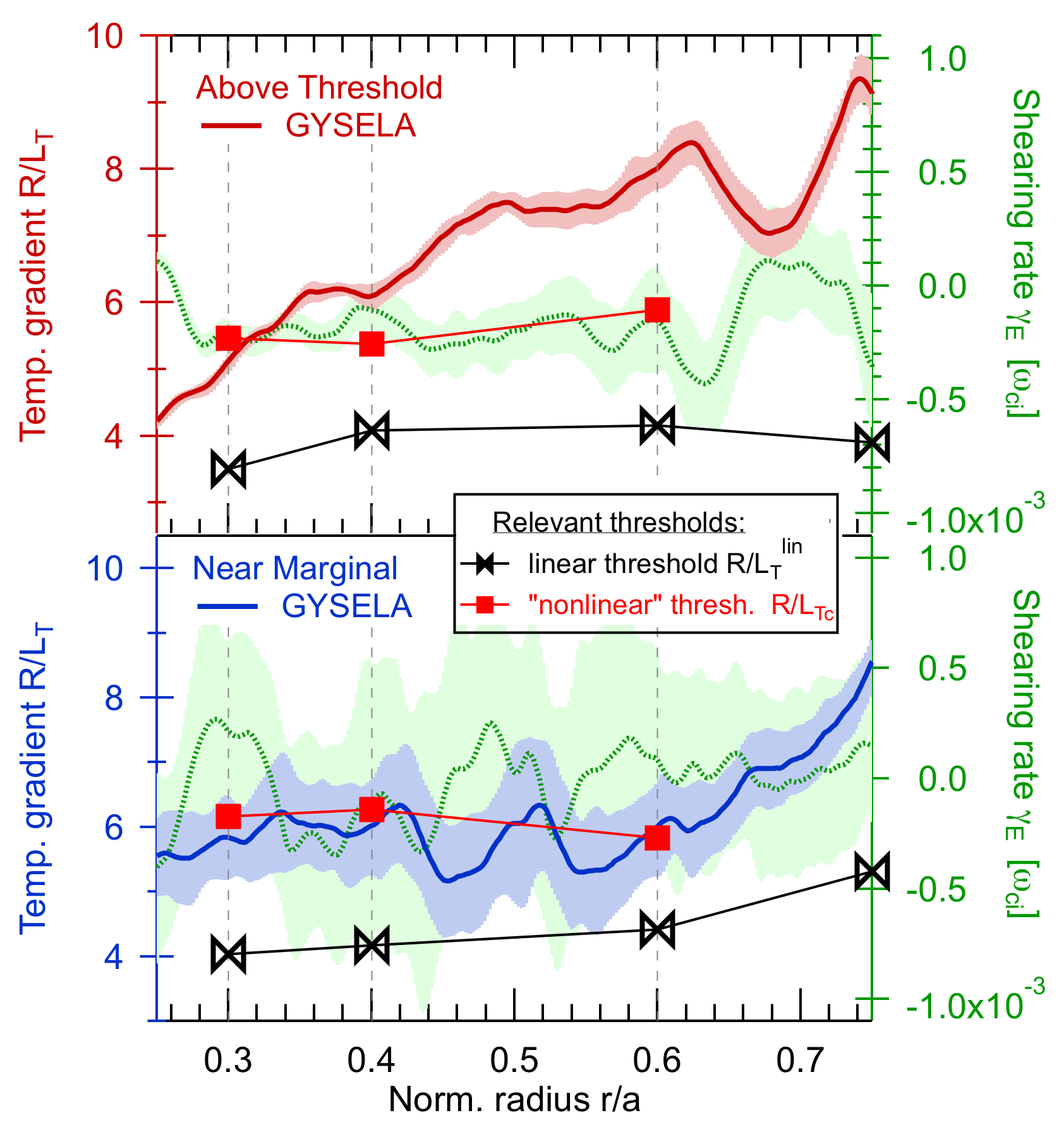}}
    \caption{
        Radial profiles of normalised temperature gradients (blue and red, left axis)
        and zonal flow shear (green, right axis) for the 'above threshold' (top) and 'near marginal' (bottom) cases. Shaded areas represent temporal standard deviation; black hourglass symbols 
($\Join$) linear instability thresholds $R/L_T^{\text{lin}}$ at vanishing $\exb$ shear; red squares (${\color{red}\blacksquare}$) nonlinear thresholds $R/L_{T c}$ estimated by \gkw\ in the local limit, including the \gys\ $\exb$ shear.
        \label{fig:RonLT}
    }
\end{figure}

\begin{table}[!b]
\begin{center}
\begin{tabular}{|c|c|c|c|c|c|}
    \hline\hline
    Case & $\rho_{\star,50}$ 
    & \newstuff{$R/a$} & $\nu_{\star,50}$ & $(q_{50}\,;\,q_{95})$ & $\tau=T_i/T_e$  \\
    \hline
    near marginal & $1/250$ 
    & $3.2$ & $0.24$ & $(1.4 \,;\, 4.0)$ & $1<\tau<1.3$ \\\hline
    above thresh. & $1/350$ 
    & $6$  & $0.02$ & $( 1.7 \,;\, 2.8)$ & $0.9<\tau<1$  \\    
    \hline\hline
\end{tabular}
\end{center}
\caption{Main plasma parameters in considered cases. Subscripts $_{50}$ and $_{95}$ respectively denote parameter values estimated at locations $r/a=0.5$ and $r/a=0.95$.}
\label{tab_params}
\end{table}
{\gys} resolves ion Larmor radius scale turbulence and collisional transport in global tokamak geometry with no scale separation between equilibrium and fluctuations. A centrally peaked heat source drives a deuterium plasma out of equilibrium, which converges towards a steady temperature profile $T$ on energy confinement times. Two paradigmatic simulation regimes are considered in the electrostatic regime with Boltzmann electrons. Both cases are run in the so-called ``local limit'' \cite{Lin_2002}, at $\rho_\ast = \rho_i/a \leq 1/250$, where comparison to local \gkw\ is fair. "Local" simulations exploit to numerical advantage the assumption of a scale separation between equilibrium and fluctuations and solve a limited subset of the whole plasma volume which twists around the torus due to the magnetic shear of the background magnetic equilibrium. In this approximation, local fluxes bijectively relate to local gradients. Here $\rho_i$ and $a$ are respectively the local Larmor radius and minor radius of the torus. 
\newstuff{
The time averaged (over several tens of turbulence correlation times in each case) radial profiles of normalised temperature gradients $R / L_T = - R \partial_r T / T$ and zonal flow shear are plotted in Fig.\ref{fig:RonLT} for the `Above Threshold' (top) and `Near Marginal' (bottom) cases. The shaded areas represent temporal standard deviation. The large deviation from the mean shearing rate in the near marginal case (bottom, right axis) results from the meandering of staircases, which have already been reported to play an important role in this regime of parameters \cite{Dif-Pradalier_2017}. The linear (black hourglass symbols) and nonlinear (red squares) thresholds are discussed in next section.
} 
In order to broadly span parameter regimes, main plasma parameters vary significantly between cases, as illustrated in Table \ref{tab_params}. 

For each set of local values of the \gys\ parameters, two thresholds are to be distinguished: linear $R/L_T^{\text{lin}}$ represents the normalised temperature gradient above which an unstable mode grows, at vanishing $\exb$ shear. The inclusion of self-generated flow shear --nonlinear in essence-- introduces a second threshold $R/L_{Tc}$, indicative of the nonlinear saturation of turbulence \cite{Dimits_2000}, in particular by zonal flows. Practically, $R/L_{Tc}$ is computed in local GD frameworks as the minimum {local} temperature gradient which provides a non-vanishing nonlinear heat flux. It is estimated with \gkw\ through a series of increasing $R/L_T$ nonlinear computations, whilst imposing local $\exb$ shear values {from flux-driven} \gys. We call ``above threshold'' the regime such that $R/L_T^{\text{lin}}< R/L_{Tc} \ll  R/L_T$, characterised by a static, smooth zonal mean shear $\gamma_E = \partial_r^2 \langle \phi \rangle/B$ and subdominant zonal fluctuations. The flux-surface average of the electric potential is $\langle \phi \rangle$, $\tilde{\phi}$ are its fluctuations and $R$ is the tokamak major radius. We call ``near marginal'' regimes such that $R/L_T^{\text{lin}} < R/L_T \lessapprox R/L_{Tc}$. Proximity from below to $R/L_{Tc}$ is a hallmark of near marginality, featuring a well-defined staircase pattern of flows and associated temperature corrugations which meanders within the time interval --hence the large shear deviation-- and significant avalanching activity \cite{Dif-Pradalier_2017}. 

\section{Kubo numbers of order unity}
QLT is valid \cite{Diamond_2010} in the low Kubo number limit $K=\tau_{\text{jump}}/\tau_{\text{int}}< 1$, ratio of a jumping time $\tau_{\text{jump}}$ of particles from one turbulent eddy to the next over a nonlinear eddy-particle interaction time $\tau_{\text{int}}$. Kubo numbers are estimated from the \gys\ flux-driven data in various ways, summarised in Tab.\ref{tab_kubo}. With analogy to incompressible fluids, particles trapped in turbulent convective cells explore eddies in a typical turn-over time given by the local vorticity $\tau_{\text{int}}^{\text{trap}} \sim B / \langle | \nabla_{\bot}^2 \tilde{\phi} |^2 \rangle^{1/2}$. Whilst they undergo this vortical motion they also drift in about $\tau_{\text{jump}}^\star \sim L_\theta (eB/\nabla T)$ from one turbulent structure to the next at the typical speed of the local diamagnetic velocity. Alternatively, the slower evolution of the potential field provides a relevant correlation time $\tau_{\text{jump}}^{{\text{corr}}}$ for turbulent fluctuations, trade-off between unstable growth and nonlinear saturation. It must be computed as a Lagrangian correlation time, in the co-moving frame of the eddies to correct for the Doppler shift induced by their phase velocity. We estimate it through image registration, following the toroidal shift between 3-dimensional turbulence snapshots of $\tilde \phi$. It is compared to turbulence-driven stochastic transport times of particles which, assuming a diffusive ansatz for $\exb$ fluctuations, drift across eddies in about $\tau_{\text{int}}^{{\text{diff}},x} = L_x / \langle |\tilde v_{E,x}|^2 \rangle^{1/2}$, with $x=\{r, \theta \}$ and $L_x$ the transverse correlation lengths computed from {\gys} outputs. Interestingly, employing a Eulerian correlation time would result in severe Kubo number underestimation, locally up to factors of 25. 
\begin{table}[b]
    \begin{center}
        \begin{tabular}{|c|c|}
            \hline\hline
            \multirow{2}{*}{Particle trapping $K_{\text{trap}}$} & transverse drifts $\tau_{\text{jump}}^\star$ \\\cline{2-2}
            &  eddy turn-over $ \tau_{\text{int}}^{\text{trap}}$ \\
            \hline\hline
            \multirow{2}{*}{Random walks $K_{{\text{diff}}}^{\{r, \theta \}}$} &  Lagrang. correlation time $\tau_{\text{jump}}^{{\text{corr}}}$ \\\cline{2-2}
            & $\exb$ random walk $ \tau_{\text{int}}^{{\text{diff}},\{r, \theta \}}$ \\    
            \hline\hline
        \end{tabular}
    \end{center}
    \caption{Five typical wave--particle and turbulent times lead to three Kubo number combinations, shown in Fig.\ref{fig:kubo}.}
    \label{tab_kubo}
\end{table}
\begin{figure}[bt]
	\resizebox{\linewidth}{!}{\includegraphics{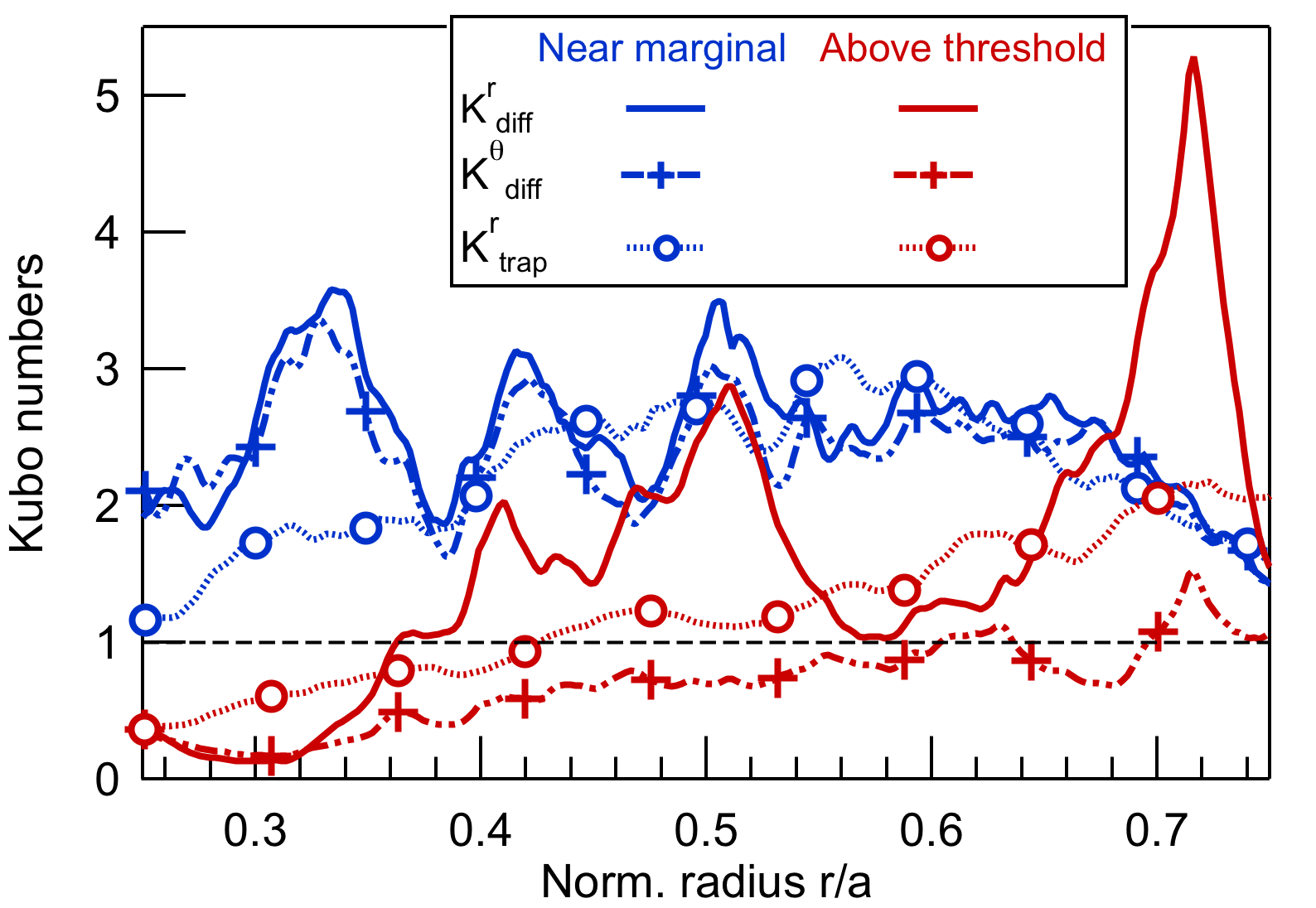}}
	\caption{
        Kubo numbers for the principal nonlinear dynamics in the problem.
        Plain and plusses: turbulent radial and poloidal $\exb$
        velocity effect during a turbulent auto-correlation time. Circles:
        trapping of	particles due to turbulent vorticity during transverse
        crossing of the	turbulent filament.
        \label{fig:kubo}
    }
\end{figure}

\begin{figure*}[!t]
	\resizebox{.93\linewidth}{!}{\includegraphics{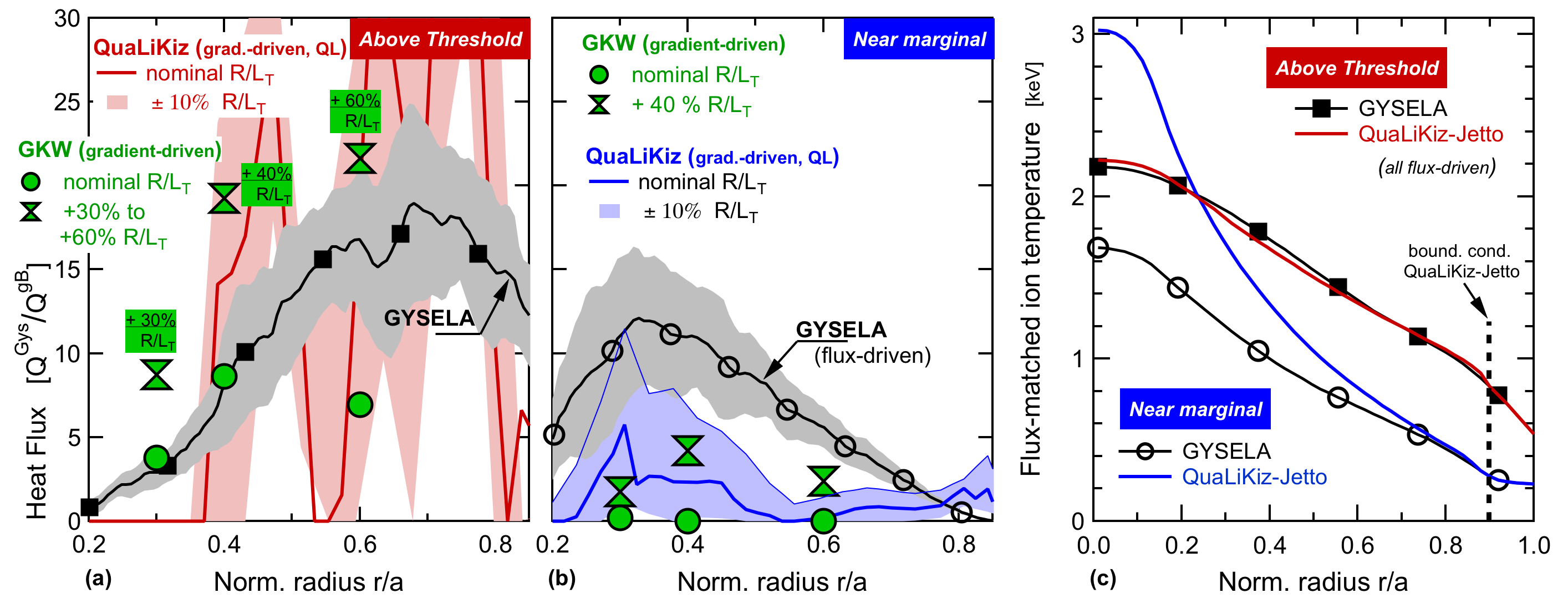}}
	\caption{
		(a) and (b): gradient-driven {\qlk} and {\gkw} heat fluxes confronted to reference flux-driven \gys\ flux levels, based on the \gys\ profiles of Fig.\ref{fig:RonLT} \newstuff{and expressed in gyro-Bohm unit}. Gray shaded areas represent a standard deviation of \gys\ heat fluxes during the considered time interval, as profiles fluctuate. Red and blue shaded areas represent the sensitivity of \qlk\ to these profile variations; the hourglass symbols that of \gkw\ to a increased input temperature gradients. The reversed approach is followed in panel (c): heat fluxes in \qlkjet\ are made to match the \gys\ reference fluxes; the unknowns are the \qlkjet\ profiles. The remarkable agreement above threshold and large over-prediction of the temperature gradient near marginality are  consistent with results in panels (a) and (b), emphasising reduction adequacy far from marginality and missing physical ingredients in its vicinity. 
		\label{fig:heat-ratio}
	}
\end{figure*}
Three Kubo numbers, combinations of the above nonlinear times are plotted in Fig.\ref{fig:kubo} for both `Above Threshold' and `Near Marginal' regimes. The various definitions for $K$, coherent, provide the following picture: (i) on the basis of order unity Kubo numbers, QLT should be marginally valid. Yet, as shown below, key assumptions at the heart of the QLT reduction remain valid throughout nonlinear evolution, which strengthens the case for quasi-linear integrated modelling. Interestingly also, (ii) consistently larger $K$ values near marginality stress the more percolative nature of transport there. Avalanching emerges as a key theme to distinguish between `Above Threshold' and `Near Marginal' regimes as they likely underpin the larger $K$ values computed near marginality. It is a likely indication that incorporating avalanching and its zonal mean flow regulation may significantly alter transport predictions and improve model behaviour near marginality. 

\section{Near marginal heat flux underprediction}
Heat fluxes are computed with {\qlk} and \gkw\ from 
\newstuff{
the \gys\ time-averaged steady-state profiles plotted in Fig.\ref{fig:RonLT} 
}
with the same degree of approximations (electrostatic \& Boltzmann electrons). All codes have different normalisations, with \gys\ e.g. being normalized to $\hat Q^{\text{Gys}} = Q/(n_0T_0 v_{Ti0})$ with $v_{Ti0} = (T_0/m_i)^{1/2}$ the ion thermal velocity. In Fig.\ref{fig:heat-ratio}--(a) and (b), fluxes from all codes are in the units of $\hat Q = \hat Q^{\text{Gys}} / \hat Q^{\text{gB}} $, with $\hat Q^{\text{gB}}= \rho_{\star,50}^2\, a/R$ and values of $\rho_{\star,50}$ and $R/a$ given in Tab.\ref{tab_params}. Consistent rescaling factors have been applied to {\qlk} and \gkw\ fluxes. 
The stabilising effect of zonal flow shear is accounted for, locally imposing the reference \gys\  shear values. Non-axisymmetric (turbulent) contributions to heat fluxes are shown in Fig.\ref{fig:heat-ratio}--(a) and (b). Turbulence spreading and profile corrugations, inherent to flux-driven complexity are absent or hindered in QL or GD approaches. A fair comparison thus requests that the \gys\ reference data be significantly coarse-grained before being handed over to \gkw\ and \qlk\ as input profiles (including $\exb$ shear profiles) and after computations, when comparing e.g. fluxes. Practically, \gys\ observables are time averaged well over a typical linear growth time ($\geq 30,000\, \Omega_{c i}$), at steady-state and a radial smoothing is performed over $60\,\rho_i$ to smear out visible FD specificities in the profiles ---FD turbulence correlation lengths and profile corrugations are indeed about $10\rho_i$ \cite{Dif-Pradalier_2017}. Here, $\Omega_{c i}$ denotes the ion cyclotron frequency. Sensitivity to gradient fluctuations, inherent to FD approaches is further assessed by additional scans in $R/L_T$ and $\exb$ shear within one temporal standard deviation. 

Without inclusion of $\exb$ shear stabilisation {\it[i]} \qlk\ heat fluxes are overestimated with respect to Fig.\ref{fig:heat-ratio} by over an order of magnitude (not shown here). With the inclusion of shear, {\it[ii]} at locations of low or vanishing $\exb$ shear and despite the $60\,\rho_i$ smoothing, gradient-driven (standalone) \qlk\ commonly displays [subplots (a) and (b)] variations by factors in heat fluxes from one radial position to the next whilst \gkw\ exhibits much less sensitivity to $\exb$ shear stabilisation. Interestingly, this large sensitivity of \qlk\ to shear is mitigated when called within the integrated framework of \textsc{Jetto} \cite{Romanelli_14} to allow for a flux-driven QL profile evolution, driven by a central source that mimics the one of \gys\ [subplot (c)]. In the regime above threshold {\it[iii]} reasonable agreement in computed fluxes is found across fidelity hierarchy. Conversely near marginality, {\it[iv]} despite significant smoothing, heat flux discrepancies in Fig.\ref{fig:heat-ratio}--(b) are well outside allowed gradient sensitivity and fluctuation `error' bars. Secular growth of zonal flows in \gkw\ is responsible for the observed large flux underprediction. This echoes previous observations casting concern on near marginal gradient-driven predictions \cite{Peeters_2016}. The soundness of separating fluctuations from mean is thus clearly questioned near marginality. 

The conclusions above are further confirmed when comparing FD \gys\ to the FD quasilinear framework of \qlkjet, in the same two regimes. In subplot (c), profiles from \qlkjet\ are evolved until heat fluxes match the nonlinear \gys\ reference fluxes. The figure of merit now becomes how close quasilinear profiles are at flux equilibrium with those of \gys. Remarkable {\it[v]} profile agreement above threshold echoes the agreement in fluxes, displayed in panel (a). Near marginality however, {\it[vi]} a large over-prediction of the temperature gradient is required for \qlkjet\ to carry the same flux as \gys. Differently stated, both GD nonlinear fluxes and QL fluxes (irrespective to the forcing framework) are underpredicted near marginality, confirming that a constitutive ingredient is missing in that regime, at least with Boltzmann electrons. 

\section{Axis for improvement: saturation rules} 
Flux discrepancies between \gys\ and \gkw\ likely stems from disregarding the feedback of fluctuations on an assumed fixed "equilibrium". This enforces local and single-valued flux--gradient relations, underestimates turbulence spreading, avalanching and mesoscale organisation. All of which contribute to transport, especially near marginality. Discrepancies with \qlk\ may either come from violating assumptions central to QLT --linearity of fluctuations-- or by inheriting shortcomings akin to those of \gkw, through the choice of saturation rules. 
\begin{figure}[t]
	\resizebox{\linewidth}{!}{\includegraphics{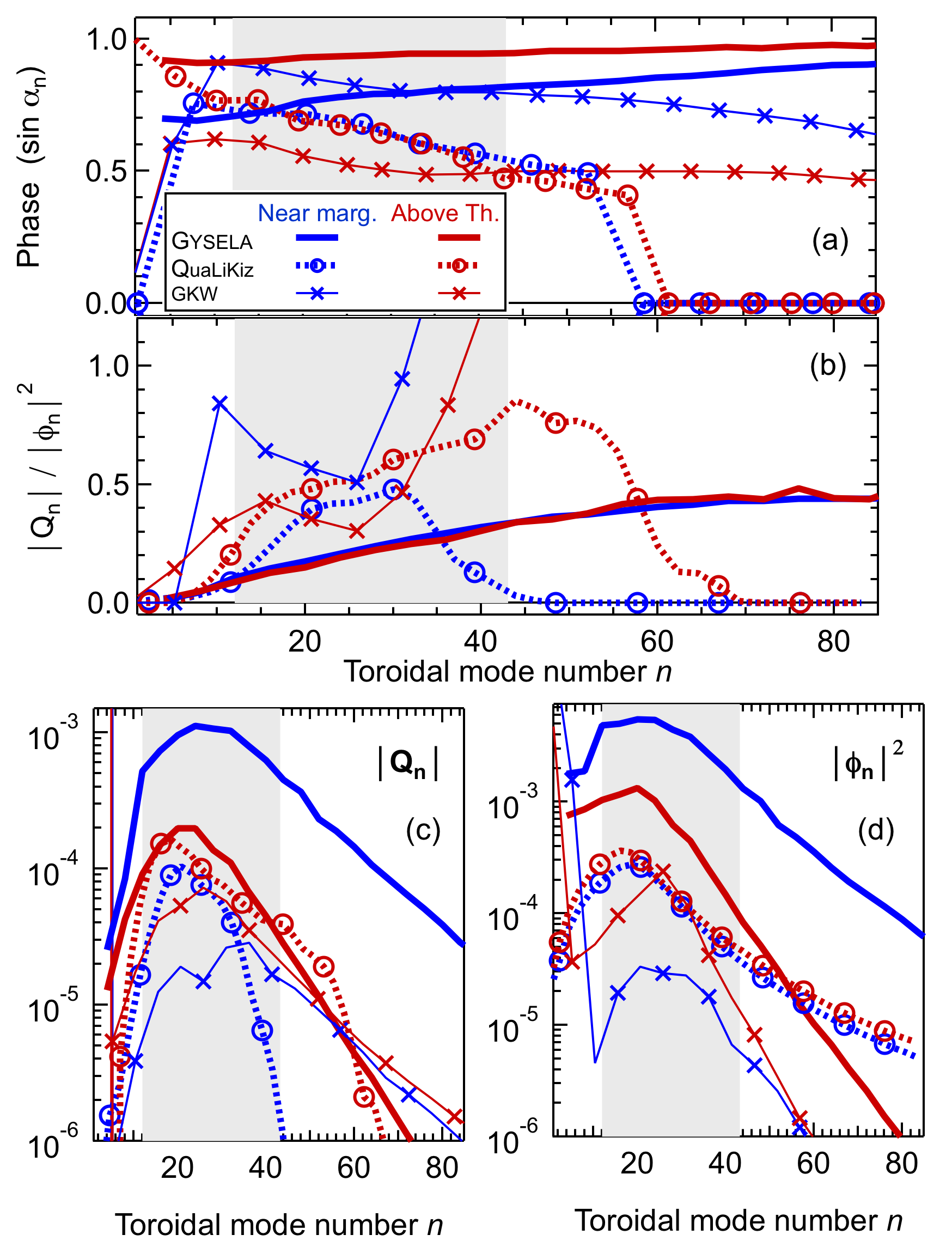}}
	\caption{
        Testing spectral distribution of key fundamental quantities: (a) linear cross-phases, (c) heat flux, (d) saturation rule and (b) their ratio, proxy to the dispersion relation from {\gys} ($\pmb -$), \gkw\ ($\times$) and {\qlk} ($\circ$), averaged over radial interval $0.3 \leqslant r / a \leqslant 0.7$. 
        \label{fig:phase-tor}
    }
\end{figure}

To disentangle these questions, we compute in Fig.\ref{fig:phase-tor}, for all three approaches, the complex argument $\alpha_n$ of the $n$-th Fourier component of the heat flux [panel (a)] --a proxy for the linear cross-phase between electric potential and pressure fluctuations. They depend on the toroidal wave number $n$ labelling each eigenmode which is related to the normalised poloidal wave vector $k_\theta\rho_i$ through: $k_\theta\rho_i = (a/r) \,n\,q\, \rho_\star$. In the present cases, the $n=25$ toroidal wave number corresponds to, at mid-radius $a/r=2$: $k_\theta\rho_i = 25*1.4*2/250 = 0.28$ and $k_\theta\rho_i = 25 *1.7 * 2/350 \approx 0.24$ for the Near Marginal and Above Threshold cases, respectively. In {\gys}, $Q_n^{\text{Gys}} = -i |Q_n| \mathe^{i\alpha_n}= \int \hat{v}_{E, n}^r \frac{2}{3} \left( \hat{P}^{\ast}_{| |, n}/2 + \hat{P}^{\ast}_{\bot, n} \right)  \mathd \theta $, where parallel and transverse components of the pressure and the radial component of the $\exb$ drift are computed from 3D output data. The $n$-Fourier components of the heat flux $|Q_n|$ and of the squared potential $|\phi_n|^2$, i.e. the saturation rule are respectively shown in panels (c) and (d). Panel (b) displays their ratio, a proxy for the dispersion relation, sometimes called ‘QL flux integrals’.

Clearly, linear cross-phases display reasonable agreement, within factors of 2 and across all regimes. They are not responsible for the flux discrepancies. Factors of disagreement --and avenues of improvement for QL modeling-- are essentially twofold: {\it[i]} $90\%$ of the heat flux is carried by modes $n\in\lbrack\!\lbrack 5,50 \rbrack\!\rbrack$ in the above threshold regime; all approaches provide similar conclusions. Near marginality however, flux is carried in \gys\ and \gkw\ through $n\in\lbrack\!\lbrack 5,70 \rbrack\!\rbrack$, twice the amount of active modes with respect to \qlk. Furthermore, {\it [ii]} saturation rules are clearly responsible for much of the observed flux discrepancies. In the above threshold regime, flux spectra agree well [panel (c)] --though this results from a surprising compensation: the potential, or saturation rule [panel (d)] is under-estimated, the dispersion relation [panel (b)] is over-estimated yet heat flux spectra in the above threshold regime are in reasonable agreement. No such compensation occurs near marginality; there severe under-estimation of the potential spectrum is clearly responsible for the under-prediction of fluxes.

\section{Outlook} 
The success of reduced models especially hinges on the reproduction of nonlinear gyrokinetic fluxes \cite{Citrin_2017}. Flux under-prediction in the dynamically important regime of near marginal stability is thus a  matter of importance. At the heart of this paper lies the fact that FD and GD models provide significantly different flux predictions with Bolzmann electrons close to marginal stability $R/L_{Tc}$, which underpins basic discrepancies in how nonlinear saturation of turbulence is modelled. This should foster renewed interest in ways to complete QL or GD models near marginality. The robustness of linear features \cite{Besse_2011} across fidelity hierarchy and across turbulent regimes is encouraging perspective and provides constructive directions whereby reduced models could be improved.

Discrepancies between \gys\ and \gkw\ are strong indications that turbulence spreading and mesoscale patterning are in fact central to accurate transport predictions near marginality. Larger Kubo numbers near marginality, as shown in Fig.\ref{fig:kubo} reinforce this point, which is also differently stressed by recent works in the plasma edge \cite{Singh_PoP20, DifPradalier_encours22}. The fact that \qlk\ behaves better near marginality than \gkw\ is likely because the QL closure does not include the long-lived zonal flows that quench transport near marginality in GD approaches \cite{Peeters_2016}. In \qlk, stable modes are neglected and the turbulent intensity spectrum --chosen as a double power law in $k_\theta$ for unstable modes-- is fitted \cite{Casati_2009, Citrin_2017} onto databases of GD nonlinear computations, similar to those presented here with \gkw. With this procedure, QL models inherit the shortcomings of the primitive GD models onto which they are adjusted. The near marginal transport shortfall in \qlk\ indeed largely comes from issues with the QL closure, i.e. the choice of saturation rule. 

This provides directions for improvement. In physical terms, near marginal regimes require description of transport below or at linear stability and coupling to modes presently predicted as stable in \qlk. It also requires to model the self-advection (spreading) of turbulent domains and the possibility of non-monotonic flux-gradient relations. Several routes can be explored. In the spirit of current frameworks, QL models could be trained on near marginal flux-driven databases such as provided by the likes of \gys. This would likely lead to close QL models with regime-dependent turbulent intensity spectra. Alternatively to present closures, QL models could also be coupled to dynamic equations for the turbulence intensity, e.g. in the form of reaction--diffusion \cite{Heinonen_PoP19} or $k-\epsilon$ equations \cite{Baschetti_NF21}, enriching accessible nonlinear dynamics. 




Ongoing studies are concerned with further characterising our “Above Threshold” and “Near Marginal” regimes \cite{Garbet_PoP07} when kinetic features of electron dynamics are present. This is important to assess relevance for ITER extrapolations. Electron dynamics is known to locally modify turbulence organisation near low order rational surfaces \cite{Dominski_2012}. Interestingly however, key features of near marginal turbulence with Boltzmann electrons (flow patterning, shear effectiveness and staircase organisation) which are central here to the near marginal regime robustly endure in kinetic electron regimes \cite{Rath_PoP21}. The present work provides a framework of understanding; it will be expanded to various parameter regimes. 

\section{Acknowledgements}
The authors acknowledge stimulating discussions with P.H. Diamond and participants at the 2019 and 2021 "Festival de Th\'eorie" in Aix-en-Provence. This work has been carried out within the framework of the EUROfusion Consortium and was supported by the EUROfusion Theory and Advanced Simulation Coordination (E-TASC) initiative under the TSVV (Theory, Simulation, Verification and Validation) “L-H transition and pedestal physics” project (2019–2020) and TSVV “Plasma Particle/Heat Exhaust: Gyrokinetic/Kinetic Edge Codes” (2021–2025). It has received funding from the Euratom research and training programme 2014-2018 and 2019-2020 under grant agreement No 633053. The authors gratefully acknowledge funding from the European Commission Horizon 2020 research and innovation program under Grant Agreement No. 824158 (EoCoE-II). The views and opinions expressed herein do not necessarily reflect those of the European Commission. This research was supported in part by the National Science Foundation under Grant No. NSF PHY-1748958. We acknowledge PRACE for awarding us access to Joliot-Curie at GENCI@CEA, France and MareNostrum at Barcelona Supercomputing Center (BSC), Spain.


\begin{thebibliography}{40}%
\makeatletter
\providecommand \@ifxundefined [1]{%
 \@ifx{#1\undefined}
}%
\providecommand \@ifnum [1]{%
 \ifnum #1\expandafter \@firstoftwo
 \else \expandafter \@secondoftwo
 \fi
}%
\providecommand \@ifx [1]{%
 \ifx #1\expandafter \@firstoftwo
 \else \expandafter \@secondoftwo
 \fi
}%
\providecommand \natexlab [1]{#1}%
\providecommand \enquote  [1]{``#1''}%
\providecommand \bibnamefont  [1]{#1}%
\providecommand \bibfnamefont [1]{#1}%
\providecommand \citenamefont [1]{#1}%
\providecommand \href@noop [0]{\@secondoftwo}%
\providecommand \href [0]{\begingroup \@sanitize@url \@href}%
\providecommand \@href[1]{\@@startlink{#1}\@@href}%
\providecommand \@@href[1]{\endgroup#1\@@endlink}%
\providecommand \@sanitize@url [0]{\catcode `\\12\catcode `\$12\catcode
  `\&12\catcode `\#12\catcode `\^12\catcode `\_12\catcode `\%12\relax}%
\providecommand \@@startlink[1]{}%
\providecommand \@@endlink[0]{}%
\providecommand \url  [0]{\begingroup\@sanitize@url \@url }%
\providecommand \@url [1]{\endgroup\@href {#1}{\urlprefix }}%
\providecommand \urlprefix  [0]{URL }%
\providecommand \Eprint [0]{\href }%
\providecommand \doibase [0]{https://doi.org/}%
\providecommand \selectlanguage [0]{\@gobble}%
\providecommand \bibinfo  [0]{\@secondoftwo}%
\providecommand \bibfield  [0]{\@secondoftwo}%
\providecommand \translation [1]{[#1]}%
\providecommand \BibitemOpen [0]{}%
\providecommand \bibitemStop [0]{}%
\providecommand \bibitemNoStop [0]{.\EOS\space}%
\providecommand \EOS [0]{\spacefactor3000\relax}%
\providecommand \BibitemShut  [1]{\csname bibitem#1\endcsname}%
\let\auto@bib@innerbib\@empty

\bibitem [{\citenamefont {Diamond}\ and\ \citenamefont
  {Hahm}(1995)}]{Diamond_1995}%
  \BibitemOpen
  \bibfield  {author} {\bibinfo {author} {\bibfnamefont {P.~H.}\ \bibnamefont
  {Diamond}}\ and\ \bibinfo {author} {\bibfnamefont {T.~S.}\ \bibnamefont
  {Hahm}},\ }\bibfield  {title} {\bibinfo {title} {On the dynamics of turbulent
  transport near marginal stability},\ }\href
  {https://doi.org/10.1063/1.871063} {\bibfield  {journal} {\bibinfo  {journal}
  {Physics of Plasmas}\ }\textbf {\bibinfo {volume} {2}},\ \bibinfo {pages}
  {3640} (\bibinfo {year} {1995})}\BibitemShut {NoStop}%
\bibitem [{\citenamefont {Thirring}(1970)}]{Thirring_1970}%
  \BibitemOpen
  \bibfield  {author} {\bibinfo {author} {\bibfnamefont {W.}~\bibnamefont
  {Thirring}},\ }\bibfield  {title} {\bibinfo {title} {Systems with negative
  specific heat},\ }\href {https://doi.org/10.1007/BF01403177} {\bibfield
  {journal} {\bibinfo  {journal} {Zeitschrift f{\"u}r Physik A Hadrons and
  nuclei}\ }\textbf {\bibinfo {volume} {235}},\ \bibinfo {pages} {339}
  (\bibinfo {year} {1970})}\BibitemShut {NoStop}%
\bibitem [{\citenamefont {Ellis}\ \emph {et~al.}(2000)\citenamefont {Ellis},
  \citenamefont {Haven},\ and\ \citenamefont {Turkington}}]{Ellis_2000}%
  \BibitemOpen
  \bibfield  {author} {\bibinfo {author} {\bibfnamefont {R.~S.}\ \bibnamefont
  {Ellis}}, \bibinfo {author} {\bibfnamefont {K.}~\bibnamefont {Haven}},\ and\
  \bibinfo {author} {\bibfnamefont {B.}~\bibnamefont {Turkington}},\ }\bibfield
   {title} {{\selectlanguage {English}\bibinfo {title} {Large deviation
  principles and complete equivalence and nonequivalence results for pure and
  mixed ensembles}},\ }\href {https://doi.org/10.1023/A:1026446225804}
  {\bibfield  {journal} {\bibinfo  {journal} {Journal of Statistical Physics}\
  }\textbf {\bibinfo {volume} {101}},\ \bibinfo {pages} {999} (\bibinfo {year}
  {2000})}\BibitemShut {NoStop}%
\bibitem [{\citenamefont {Saint-Michel}\ \emph {et~al.}(2013)\citenamefont
  {Saint-Michel}, \citenamefont {Dubrulle}, \citenamefont {Mari\'e},
  \citenamefont {Ravelet},\ and\ \citenamefont {Daviaud}}]{Saint-Michel_2013}%
  \BibitemOpen
  \bibfield  {author} {\bibinfo {author} {\bibfnamefont {B.}~\bibnamefont
  {Saint-Michel}}, \bibinfo {author} {\bibfnamefont {B.}~\bibnamefont
  {Dubrulle}}, \bibinfo {author} {\bibfnamefont {L.}~\bibnamefont {Mari\'e}},
  \bibinfo {author} {\bibfnamefont {F.}~\bibnamefont {Ravelet}},\ and\ \bibinfo
  {author} {\bibfnamefont {F.}~\bibnamefont {Daviaud}},\ }\bibfield  {title}
  {\bibinfo {title} {Evidence for forcing-dependent steady states in a
  turbulent swirling flow},\ }\href
  {https://doi.org/10.1103/PhysRevLett.111.234502} {\bibfield  {journal}
  {\bibinfo  {journal} {Phys. Rev. Lett.}\ }\textbf {\bibinfo {volume} {111}},\
  \bibinfo {pages} {234502} (\bibinfo {year} {2013})}\BibitemShut {NoStop}%
\bibitem [{\citenamefont {Newman}\ \emph {et~al.}(1996)\citenamefont {Newman},
  \citenamefont {Carreras}, \citenamefont {Diamond},\ and\ \citenamefont
  {Hahm}}]{Newman_1996}%
  \BibitemOpen
  \bibfield  {author} {\bibinfo {author} {\bibfnamefont {D.~E.}\ \bibnamefont
  {Newman}}, \bibinfo {author} {\bibfnamefont {B.~A.}\ \bibnamefont
  {Carreras}}, \bibinfo {author} {\bibfnamefont {P.~H.}\ \bibnamefont
  {Diamond}},\ and\ \bibinfo {author} {\bibfnamefont {T.~S.}\ \bibnamefont
  {Hahm}},\ }\bibfield  {title} {\bibinfo {title} {The dynamics of marginality
  and self-organized criticality as a paradigm for turbulent transport},\
  }\href {https://doi.org/10.1063/1.871681} {\bibfield  {journal} {\bibinfo
  {journal} {Physics of Plasmas}\ }\textbf {\bibinfo {volume} {3}},\ \bibinfo
  {pages} {1858} (\bibinfo {year} {1996})}\BibitemShut {NoStop}%
\bibitem [{\citenamefont {Garbet}\ and\ \citenamefont
  {Waltz}(1998)}]{Garbet_1998}%
  \BibitemOpen
  \bibfield  {author} {\bibinfo {author} {\bibfnamefont {X.}~\bibnamefont
  {Garbet}}\ and\ \bibinfo {author} {\bibfnamefont {R.~E.}\ \bibnamefont
  {Waltz}},\ }\bibfield  {title} {\bibinfo {title} {Heat flux driven ion
  turbulence},\ }\href {https://doi.org/10.1063/1.873003} {\bibfield  {journal}
  {\bibinfo  {journal} {Physics of Plasmas}\ }\textbf {\bibinfo {volume} {5}},\
  \bibinfo {pages} {2836} (\bibinfo {year} {1998})}\BibitemShut {NoStop}%
\bibitem [{\citenamefont {Beyer}\ \emph {et~al.}(2000)\citenamefont {Beyer},
  \citenamefont {Benkadda}, \citenamefont {Garbet},\ and\ \citenamefont
  {Diamond}}]{Beyer_2000}%
  \BibitemOpen
  \bibfield  {author} {\bibinfo {author} {\bibfnamefont {P.}~\bibnamefont
  {Beyer}}, \bibinfo {author} {\bibfnamefont {S.}~\bibnamefont {Benkadda}},
  \bibinfo {author} {\bibfnamefont {X.}~\bibnamefont {Garbet}},\ and\ \bibinfo
  {author} {\bibfnamefont {P.~H.}\ \bibnamefont {Diamond}},\ }\bibfield
  {title} {\bibinfo {title} {Nondiffusive transport in tokamaks:
  Three-dimensional structure of bursts and the role of zonal flows},\ }\href
  {https://doi.org/10.1103/physrevlett.85.4892} {\bibfield  {journal} {\bibinfo
   {journal} {Physical Review Letters}\ }\textbf {\bibinfo {volume} {85}},\
  \bibinfo {pages} {4892} (\bibinfo {year} {2000})}\BibitemShut {NoStop}%
\bibitem [{\citenamefont {McMillan}\ \emph {et~al.}(2009)\citenamefont
  {McMillan}, \citenamefont {Jolliet}, \citenamefont {Tran}, \citenamefont
  {Villard}, \citenamefont {Bottino},\ and\ \citenamefont
  {Angelino}}]{McMillan_2009}%
  \BibitemOpen
  \bibfield  {author} {\bibinfo {author} {\bibfnamefont {B.~F.}\ \bibnamefont
  {McMillan}}, \bibinfo {author} {\bibfnamefont {S.}~\bibnamefont {Jolliet}},
  \bibinfo {author} {\bibfnamefont {T.~M.}\ \bibnamefont {Tran}}, \bibinfo
  {author} {\bibfnamefont {L.}~\bibnamefont {Villard}}, \bibinfo {author}
  {\bibfnamefont {A.}~\bibnamefont {Bottino}},\ and\ \bibinfo {author}
  {\bibfnamefont {P.}~\bibnamefont {Angelino}},\ }\bibfield  {title} {\bibinfo
  {title} {Avalanchelike bursts in global gyrokinetic simulations},\ }\href
  {https://doi.org/10.1063/1.3079076} {\bibfield  {journal} {\bibinfo
  {journal} {Physics of Plasmas}\ }\textbf {\bibinfo {volume} {16}},\ \bibinfo
  {pages} {022310} (\bibinfo {year} {2009})}\BibitemShut {NoStop}%
\bibitem [{\citenamefont {Idomura}\ \emph {et~al.}(2009)\citenamefont
  {Idomura}, \citenamefont {Urano}, \citenamefont {Aiba},\ and\ \citenamefont
  {Tokuda}}]{Idomura_2009}%
  \BibitemOpen
  \bibfield  {author} {\bibinfo {author} {\bibfnamefont {Y.}~\bibnamefont
  {Idomura}}, \bibinfo {author} {\bibfnamefont {H.}~\bibnamefont {Urano}},
  \bibinfo {author} {\bibfnamefont {N.}~\bibnamefont {Aiba}},\ and\ \bibinfo
  {author} {\bibfnamefont {S.}~\bibnamefont {Tokuda}},\ }\bibfield  {title}
  {\bibinfo {title} {Study of ion turbulent transport and profile formations
  using global gyrokinetic full-$f$ {Vlasov} simulation},\ }\href
  {https://doi.org/10.1088/0029-5515/49/6/065029} {\bibfield  {journal}
  {\bibinfo  {journal} {Nuclear Fusion}\ }\textbf {\bibinfo {volume} {49}},\
  \bibinfo {pages} {065029} (\bibinfo {year} {2009})}\BibitemShut {NoStop}%
\bibitem [{\citenamefont {Sarazin}\ \emph {et~al.}(2010)\citenamefont
  {Sarazin}, \citenamefont {Grandgirard}, \citenamefont {Abiteboul},
  \citenamefont {Allfrey}, \citenamefont {Garbet}, \citenamefont {Ghendrih},
  \citenamefont {Latu}, \citenamefont {Strugarek},\ and\ \citenamefont
  {Dif-Pradalier}}]{Sarazin_2010}%
  \BibitemOpen
  \bibfield  {author} {\bibinfo {author} {\bibfnamefont {Y.}~\bibnamefont
  {Sarazin}}, \bibinfo {author} {\bibfnamefont {V.}~\bibnamefont
  {Grandgirard}}, \bibinfo {author} {\bibfnamefont {J.}~\bibnamefont
  {Abiteboul}}, \bibinfo {author} {\bibfnamefont {S.}~\bibnamefont {Allfrey}},
  \bibinfo {author} {\bibfnamefont {X.}~\bibnamefont {Garbet}}, \bibinfo
  {author} {\bibfnamefont {P.}~\bibnamefont {Ghendrih}}, \bibinfo {author}
  {\bibfnamefont {G.}~\bibnamefont {Latu}}, \bibinfo {author} {\bibfnamefont
  {A.}~\bibnamefont {Strugarek}},\ and\ \bibinfo {author} {\bibfnamefont
  {G.}~\bibnamefont {Dif-Pradalier}},\ }\bibfield  {title} {\bibinfo {title}
  {Large scale dynamics in flux driven gyrokinetic turbulence},\ }\href
  {https://doi.org/10.1088/0029-5515/50/5/054004} {\bibfield  {journal}
  {\bibinfo  {journal} {Nuclear Fusion}\ }\textbf {\bibinfo {volume} {50}},\
  \bibinfo {pages} {054004} (\bibinfo {year} {2010})}\BibitemShut {NoStop}%
\bibitem [{\citenamefont {Singh}\ and\ \citenamefont
  {Diamond}(2020)}]{Singh_PoP20}%
  \BibitemOpen
  \bibfield  {author} {\bibinfo {author} {\bibfnamefont {R.}~\bibnamefont
  {Singh}}\ and\ \bibinfo {author} {\bibfnamefont {P.~H.}\ \bibnamefont
  {Diamond}},\ }\bibfield  {title} {\bibinfo {title} {When does turbulence
  spreading matter?},\ }\href {https://doi.org/10.1063/1.5117835} {\bibfield
  {journal} {\bibinfo  {journal} {Physics of Plasmas}\ }\textbf {\bibinfo
  {volume} {27}},\ \bibinfo {pages} {042308} (\bibinfo {year} {2020})},\
  \Eprint {https://arxiv.org/abs/https://doi.org/10.1063/1.5117835}
  {https://doi.org/10.1063/1.5117835} \BibitemShut {NoStop}%
\bibitem [{\citenamefont {jin Kim}\ and\ \citenamefont
  {Diamond}(2003)}]{Kim_2003}%
  \BibitemOpen
  \bibfield  {author} {\bibinfo {author} {\bibfnamefont {E.}~\bibnamefont {jin
  Kim}}\ and\ \bibinfo {author} {\bibfnamefont {P.~H.}\ \bibnamefont
  {Diamond}},\ }\bibfield  {title} {\bibinfo {title} {Zonal flows and transient
  dynamics of the l-h transition},\ }\bibfield  {journal} {\bibinfo  {journal}
  {Physical Review Letters}\ }\textbf {\bibinfo {volume} {90}},\ \href
  {https://doi.org/10.1103/physrevlett.90.185006}
  {10.1103/physrevlett.90.185006} (\bibinfo {year} {2003})\BibitemShut
  {NoStop}%
\bibitem [{\citenamefont {Diamond}\ \emph {et~al.}(2005)\citenamefont
  {Diamond}, \citenamefont {Itoh}, \citenamefont {Itoh},\ and\ \citenamefont
  {Hahm}}]{Diamond_2005}%
  \BibitemOpen
  \bibfield  {author} {\bibinfo {author} {\bibfnamefont {P.~H.}\ \bibnamefont
  {Diamond}}, \bibinfo {author} {\bibfnamefont {S.-I.}\ \bibnamefont {Itoh}},
  \bibinfo {author} {\bibfnamefont {K.}~\bibnamefont {Itoh}},\ and\ \bibinfo
  {author} {\bibfnamefont {T.~S.}\ \bibnamefont {Hahm}},\ }\bibfield  {title}
  {\bibinfo {title} {Zonal flows in plasma---a review},\ }\href
  {https://doi.org/10.1088/0741-3335/47/5/r01} {\bibfield  {journal} {\bibinfo
  {journal} {Plasma Physics and Controlled Fusion}\ }\textbf {\bibinfo {volume}
  {47}},\ \bibinfo {pages} {35} (\bibinfo {year} {2005})}\BibitemShut {NoStop}%
\bibitem [{\citenamefont {Dif-Pradalier}\ \emph {et~al.}(2010)\citenamefont
  {Dif-Pradalier}, \citenamefont {Diamond}, \citenamefont {Grandgirard},
  \citenamefont {Sarazin}, \citenamefont {Abiteboul}, \citenamefont {Garbet},
  \citenamefont {Ghendrih}, \citenamefont {Strugarek}, \citenamefont {Ku},\
  and\ \citenamefont {Chang}}]{Dif-Pradalier_2010}%
  \BibitemOpen
  \bibfield  {author} {\bibinfo {author} {\bibfnamefont {G.}~\bibnamefont
  {Dif-Pradalier}}, \bibinfo {author} {\bibfnamefont {P.~H.}\ \bibnamefont
  {Diamond}}, \bibinfo {author} {\bibfnamefont {V.}~\bibnamefont
  {Grandgirard}}, \bibinfo {author} {\bibfnamefont {Y.}~\bibnamefont
  {Sarazin}}, \bibinfo {author} {\bibfnamefont {J.}~\bibnamefont {Abiteboul}},
  \bibinfo {author} {\bibfnamefont {X.}~\bibnamefont {Garbet}}, \bibinfo
  {author} {\bibfnamefont {P.}~\bibnamefont {Ghendrih}}, \bibinfo {author}
  {\bibfnamefont {A.}~\bibnamefont {Strugarek}}, \bibinfo {author}
  {\bibfnamefont {S.}~\bibnamefont {Ku}},\ and\ \bibinfo {author}
  {\bibfnamefont {C.~S.}\ \bibnamefont {Chang}},\ }\bibfield  {title} {\bibinfo
  {title} {On the validity of the local diffusive paradigm in turbulent plasma
  transport},\ }\href {https://doi.org/10.1103/PhysRevE.82.025401} {\bibfield
  {journal} {\bibinfo  {journal} {Physical Review E}\ }\textbf {\bibinfo
  {volume} {82}},\ \bibinfo {pages} {025401} (\bibinfo {year}
  {2010})}\BibitemShut {NoStop}%
\bibitem [{\citenamefont {Dif-Pradalier}\ \emph {et~al.}(2015)\citenamefont
  {Dif-Pradalier}, \citenamefont {Hornung}, \citenamefont {Ghendrih},
  \citenamefont {Sarazin}, \citenamefont {Clairet}, \citenamefont {Vermare},
  \citenamefont {Diamond}, \citenamefont {Abiteboul}, \citenamefont
  {Cartier-Michaud}, \citenamefont {Ehrlacher}, \citenamefont {Est{\`{e}}ve},
  \citenamefont {Garbet}, \citenamefont {Grandgirard}, \citenamefont
  {G{\"{u}}rcan}, \citenamefont {Hennequin}, \citenamefont {Kosuga},
  \citenamefont {Latu}, \citenamefont {Maget}, \citenamefont {Morel},
  \citenamefont {Norscini}, \citenamefont {Sabot},\ and\ \citenamefont
  {Storelli}}]{Dif-Pradalier_2015}%
  \BibitemOpen
  \bibfield  {author} {\bibinfo {author} {\bibfnamefont {G.}~\bibnamefont
  {Dif-Pradalier}}, \bibinfo {author} {\bibfnamefont {G.}~\bibnamefont
  {Hornung}}, \bibinfo {author} {\bibfnamefont {P.}~\bibnamefont {Ghendrih}},
  \bibinfo {author} {\bibfnamefont {Y.}~\bibnamefont {Sarazin}}, \bibinfo
  {author} {\bibfnamefont {F.}~\bibnamefont {Clairet}}, \bibinfo {author}
  {\bibfnamefont {L.}~\bibnamefont {Vermare}}, \bibinfo {author} {\bibfnamefont
  {P.~H.}\ \bibnamefont {Diamond}}, \bibinfo {author} {\bibfnamefont
  {J.}~\bibnamefont {Abiteboul}}, \bibinfo {author} {\bibfnamefont
  {T.}~\bibnamefont {Cartier-Michaud}}, \bibinfo {author} {\bibfnamefont
  {C.}~\bibnamefont {Ehrlacher}}, \bibinfo {author} {\bibfnamefont
  {D.}~\bibnamefont {Est{\`{e}}ve}}, \bibinfo {author} {\bibfnamefont
  {X.}~\bibnamefont {Garbet}}, \bibinfo {author} {\bibfnamefont
  {V.}~\bibnamefont {Grandgirard}}, \bibinfo {author} {\bibfnamefont
  {{\"{O}}.~D.}\ \bibnamefont {G{\"{u}}rcan}}, \bibinfo {author} {\bibfnamefont
  {P.}~\bibnamefont {Hennequin}}, \bibinfo {author} {\bibfnamefont
  {Y.}~\bibnamefont {Kosuga}}, \bibinfo {author} {\bibfnamefont
  {G.}~\bibnamefont {Latu}}, \bibinfo {author} {\bibfnamefont {P.}~\bibnamefont
  {Maget}}, \bibinfo {author} {\bibfnamefont {P.}~\bibnamefont {Morel}},
  \bibinfo {author} {\bibfnamefont {C.}~\bibnamefont {Norscini}}, \bibinfo
  {author} {\bibfnamefont {R.}~\bibnamefont {Sabot}},\ and\ \bibinfo {author}
  {\bibfnamefont {A.}~\bibnamefont {Storelli}},\ }\bibfield  {title} {\bibinfo
  {title} {Finding the elusive $\mathbf{E}\times\mathbf{B}$ staircase in
  magnetized plasmas},\ }\bibfield  {journal} {\bibinfo  {journal} {Physical
  Review Letters}\ }\textbf {\bibinfo {volume} {114}},\ \href
  {https://doi.org/10.1103/physrevlett.114.085004}
  {10.1103/physrevlett.114.085004} (\bibinfo {year} {2015})\BibitemShut
  {NoStop}%
\bibitem [{\citenamefont {Rath}\ \emph {et~al.}(2016)\citenamefont {Rath},
  \citenamefont {Peeters}, \citenamefont {Buchholz}, \citenamefont
  {Grosshauser}, \citenamefont {Migliano}, \citenamefont {Weikl},\ and\
  \citenamefont {Strintzi}}]{Rath_2016}%
  \BibitemOpen
  \bibfield  {author} {\bibinfo {author} {\bibfnamefont {F.}~\bibnamefont
  {Rath}}, \bibinfo {author} {\bibfnamefont {A.~G.}\ \bibnamefont {Peeters}},
  \bibinfo {author} {\bibfnamefont {R.}~\bibnamefont {Buchholz}}, \bibinfo
  {author} {\bibfnamefont {S.~R.}\ \bibnamefont {Grosshauser}}, \bibinfo
  {author} {\bibfnamefont {P.}~\bibnamefont {Migliano}}, \bibinfo {author}
  {\bibfnamefont {A.}~\bibnamefont {Weikl}},\ and\ \bibinfo {author}
  {\bibfnamefont {D.}~\bibnamefont {Strintzi}},\ }\bibfield  {title} {\bibinfo
  {title} {Comparison of gradient and flux driven gyro-kinetic turbulent
  transport},\ }\href {https://doi.org/10.1063/1.4952621} {\bibfield  {journal}
  {\bibinfo  {journal} {Physics of Plasmas}\ }\textbf {\bibinfo {volume}
  {23}},\ \bibinfo {pages} {052309} (\bibinfo {year} {2016})}\BibitemShut
  {NoStop}%
\bibitem [{\citenamefont {Peeters}\ \emph {et~al.}(2016)\citenamefont
  {Peeters}, \citenamefont {Rath}, \citenamefont {Buchholz}, \citenamefont
  {Camenen}, \citenamefont {Candy}, \citenamefont {Casson}, \citenamefont
  {Grosshauser}, \citenamefont {Hornsby}, \citenamefont {Strintzi},\ and\
  \citenamefont {Weikl}}]{Peeters_2016}%
  \BibitemOpen
  \bibfield  {author} {\bibinfo {author} {\bibfnamefont {A.~G.}\ \bibnamefont
  {Peeters}}, \bibinfo {author} {\bibfnamefont {F.}~\bibnamefont {Rath}},
  \bibinfo {author} {\bibfnamefont {R.}~\bibnamefont {Buchholz}}, \bibinfo
  {author} {\bibfnamefont {Y.}~\bibnamefont {Camenen}}, \bibinfo {author}
  {\bibfnamefont {J.}~\bibnamefont {Candy}}, \bibinfo {author} {\bibfnamefont
  {F.~J.}\ \bibnamefont {Casson}}, \bibinfo {author} {\bibfnamefont {S.~R.}\
  \bibnamefont {Grosshauser}}, \bibinfo {author} {\bibfnamefont {W.~A.}\
  \bibnamefont {Hornsby}}, \bibinfo {author} {\bibfnamefont {D.}~\bibnamefont
  {Strintzi}},\ and\ \bibinfo {author} {\bibfnamefont {A.}~\bibnamefont
  {Weikl}},\ }\bibfield  {title} {\bibinfo {title} {Gradient-driven flux-tube
  simulations of ion temperature gradient turbulence close to the non-linear
  threshold},\ }\href {https://doi.org/10.1063/1.4961231} {\bibfield  {journal}
  {\bibinfo  {journal} {Physics of Plasmas}\ }\textbf {\bibinfo {volume}
  {23}},\ \bibinfo {pages} {082517} (\bibinfo {year} {2016})}\BibitemShut
  {NoStop}%
\bibitem [{\citenamefont {Dif-Pradalier}\ \emph {et~al.}(2017)\citenamefont
  {Dif-Pradalier}, \citenamefont {Hornung}, \citenamefont {Garbet},
  \citenamefont {Ghendrih}, \citenamefont {Grandgirard}, \citenamefont {Latu},\
  and\ \citenamefont {Sarazin}}]{Dif-Pradalier_2017}%
  \BibitemOpen
  \bibfield  {author} {\bibinfo {author} {\bibfnamefont {G.}~\bibnamefont
  {Dif-Pradalier}}, \bibinfo {author} {\bibfnamefont {G.}~\bibnamefont
  {Hornung}}, \bibinfo {author} {\bibfnamefont {X.}~\bibnamefont {Garbet}},
  \bibinfo {author} {\bibfnamefont {P.}~\bibnamefont {Ghendrih}}, \bibinfo
  {author} {\bibfnamefont {V.}~\bibnamefont {Grandgirard}}, \bibinfo {author}
  {\bibfnamefont {G.}~\bibnamefont {Latu}},\ and\ \bibinfo {author}
  {\bibfnamefont {Y.}~\bibnamefont {Sarazin}},\ }\bibfield  {title} {\bibinfo
  {title} {The {$\mathbf{E}\times\mathbf{B}$} staircase of magnetised
  plasmas},\ }\href {https://doi.org/10.1088/1741-4326/aa6873} {\bibfield
  {journal} {\bibinfo  {journal} {Nuclear Fusion}\ }\textbf {\bibinfo {volume}
  {57}},\ \bibinfo {pages} {066026} (\bibinfo {year} {2017})}\BibitemShut
  {NoStop}%
\bibitem [{\citenamefont {Ashourvan}\ \emph {et~al.}(2019)\citenamefont
  {Ashourvan}, \citenamefont {Nazikian}, \citenamefont {Belli}, \citenamefont
  {Candy}, \citenamefont {Eldon}, \citenamefont {Grierson}, \citenamefont
  {Guttenfelder}, \citenamefont {Haskey}, \citenamefont {Lasnier},
  \citenamefont {McKee},\ and\ \citenamefont {Petty}}]{Ashourvan_2019}%
  \BibitemOpen
  \bibfield  {author} {\bibinfo {author} {\bibfnamefont {A.}~\bibnamefont
  {Ashourvan}}, \bibinfo {author} {\bibfnamefont {R.}~\bibnamefont {Nazikian}},
  \bibinfo {author} {\bibfnamefont {E.}~\bibnamefont {Belli}}, \bibinfo
  {author} {\bibfnamefont {J.}~\bibnamefont {Candy}}, \bibinfo {author}
  {\bibfnamefont {D.}~\bibnamefont {Eldon}}, \bibinfo {author} {\bibfnamefont
  {B.~A.}\ \bibnamefont {Grierson}}, \bibinfo {author} {\bibfnamefont
  {W.}~\bibnamefont {Guttenfelder}}, \bibinfo {author} {\bibfnamefont {S.~R.}\
  \bibnamefont {Haskey}}, \bibinfo {author} {\bibfnamefont {C.}~\bibnamefont
  {Lasnier}}, \bibinfo {author} {\bibfnamefont {G.~R.}\ \bibnamefont {McKee}},\
  and\ \bibinfo {author} {\bibfnamefont {C.~C.}\ \bibnamefont {Petty}},\
  }\bibfield  {title} {\bibinfo {title} {Formation of a high pressure staircase
  pedestal with suppressed edge localized modes in the diii-d tokamak},\ }\href
  {https://doi.org/10.1103/PhysRevLett.123.115001} {\bibfield  {journal}
  {\bibinfo  {journal} {Phys. Rev. Lett.}\ }\textbf {\bibinfo {volume} {123}},\
  \bibinfo {pages} {115001} (\bibinfo {year} {2019})}\BibitemShut {NoStop}%
\bibitem [{\citenamefont {Nakata}\ and\ \citenamefont
  {Idomura}(2013)}]{Nakata_2013}%
  \BibitemOpen
  \bibfield  {author} {\bibinfo {author} {\bibfnamefont {M.}~\bibnamefont
  {Nakata}}\ and\ \bibinfo {author} {\bibfnamefont {Y.}~\bibnamefont
  {Idomura}},\ }\bibfield  {title} {\bibinfo {title} {Plasma size and
  collisionality scaling of ion-temperature-gradient-driven turbulence},\
  }\href {https://doi.org/10.1088/0029-5515/53/11/113039} {\bibfield  {journal}
  {\bibinfo  {journal} {Nuclear Fusion}\ }\textbf {\bibinfo {volume} {53}},\
  \bibinfo {pages} {113039} (\bibinfo {year} {2013})}\BibitemShut {NoStop}%
\bibitem [{\citenamefont {G.~Laval}\ and\ \citenamefont
  {Adam}(2018)}]{Laval_2018}%
  \BibitemOpen
  \bibfield  {author} {\bibinfo {author} {\bibfnamefont {D.~P.}\ \bibnamefont
  {G.~Laval}}\ and\ \bibinfo {author} {\bibfnamefont {J.-C.}\ \bibnamefont
  {Adam}},\ }\bibfield  {title} {\bibinfo {title} {Wave-particle and wave-wave
  interactions in hot plasmas: a french historical point of view},\ }\href
  {https://doi.org/10.1140/epjh/e2016-70050-2} {\bibfield  {journal} {\bibinfo
  {journal} {The European Physical Journal H}\ }\textbf {\bibinfo {volume}
  {43}},\ \bibinfo {pages} {421} (\bibinfo {year} {2018})}\BibitemShut
  {NoStop}%
\bibitem [{\citenamefont {A.A.~Vedenov}\ and\ \citenamefont
  {Sagdeev}(1962)}]{Vedenov_1962}%
  \BibitemOpen
  \bibfield  {author} {\bibinfo {author} {\bibfnamefont {E.~V.}\ \bibnamefont
  {A.A.~Vedenov}}\ and\ \bibinfo {author} {\bibfnamefont {R.}~\bibnamefont
  {Sagdeev}},\ }\bibfield  {title} {\bibinfo {title} {Quasilinear theory of
  plasma oscillations},\ }\href@noop {} {\bibfield  {journal} {\bibinfo
  {journal} {Nuclear Fusion Supplement}\ }\textbf {\bibinfo {volume} {2}},\
  \bibinfo {pages} {465} (\bibinfo {year} {1962})}\BibitemShut {NoStop}%
\bibitem [{\citenamefont {Drummond}\ and\ \citenamefont
  {Pines}(1962)}]{Drummond_1962}%
  \BibitemOpen
  \bibfield  {author} {\bibinfo {author} {\bibfnamefont {W.}~\bibnamefont
  {Drummond}}\ and\ \bibinfo {author} {\bibfnamefont {D.}~\bibnamefont
  {Pines}},\ }\bibfield  {title} {\bibinfo {title} {Nonlinear stability of
  plasma oscillations},\ }\href@noop {} {\bibfield  {journal} {\bibinfo
  {journal} {Nuclear Fusion Supplement}\ }\textbf {\bibinfo {volume} {3}},\
  \bibinfo {pages} {1049} (\bibinfo {year} {1962})}\BibitemShut {NoStop}%
\bibitem [{\citenamefont {Grandgirard}\ \emph {et~al.}(2016)\citenamefont
  {Grandgirard}, \citenamefont {Abiteboul}, \citenamefont {Bigot},
  \citenamefont {Cartier-Michaud}, \citenamefont {Crouseilles}, \citenamefont
  {Dif-Pradalier}, \citenamefont {Ehrlacher}, \citenamefont {Esteve},
  \citenamefont {Garbet}, \citenamefont {Ghendrih}, \citenamefont {Latu},
  \citenamefont {Mehrenberger}, \citenamefont {Norscini}, \citenamefont
  {Passeron}, \citenamefont {Rozar}, \citenamefont {Sarazin}, \citenamefont
  {Sonnendrücker}, \citenamefont {Strugarek},\ and\ \citenamefont
  {Zarzoso}}]{Grandgirard_2016}%
  \BibitemOpen
  \bibfield  {author} {\bibinfo {author} {\bibfnamefont {V.}~\bibnamefont
  {Grandgirard}}, \bibinfo {author} {\bibfnamefont {J.}~\bibnamefont
  {Abiteboul}}, \bibinfo {author} {\bibfnamefont {J.}~\bibnamefont {Bigot}},
  \bibinfo {author} {\bibfnamefont {T.}~\bibnamefont {Cartier-Michaud}},
  \bibinfo {author} {\bibfnamefont {N.}~\bibnamefont {Crouseilles}}, \bibinfo
  {author} {\bibfnamefont {G.}~\bibnamefont {Dif-Pradalier}}, \bibinfo {author}
  {\bibfnamefont {C.}~\bibnamefont {Ehrlacher}}, \bibinfo {author}
  {\bibfnamefont {D.}~\bibnamefont {Esteve}}, \bibinfo {author} {\bibfnamefont
  {X.}~\bibnamefont {Garbet}}, \bibinfo {author} {\bibfnamefont
  {P.}~\bibnamefont {Ghendrih}}, \bibinfo {author} {\bibfnamefont
  {G.}~\bibnamefont {Latu}}, \bibinfo {author} {\bibfnamefont {M.}~\bibnamefont
  {Mehrenberger}}, \bibinfo {author} {\bibfnamefont {C.}~\bibnamefont
  {Norscini}}, \bibinfo {author} {\bibfnamefont {C.}~\bibnamefont {Passeron}},
  \bibinfo {author} {\bibfnamefont {F.}~\bibnamefont {Rozar}}, \bibinfo
  {author} {\bibfnamefont {Y.}~\bibnamefont {Sarazin}}, \bibinfo {author}
  {\bibfnamefont {E.}~\bibnamefont {Sonnendrücker}}, \bibinfo {author}
  {\bibfnamefont {A.}~\bibnamefont {Strugarek}},\ and\ \bibinfo {author}
  {\bibfnamefont {D.}~\bibnamefont {Zarzoso}},\ }\bibfield  {title} {\bibinfo
  {title} {A {5D} gyrokinetic full-$f$ global semi-lagrangian code for
  flux-driven ion turbulence simulations},\ }\href
  {https://doi.org/10.1016/j.cpc.2016.05.007} {\bibfield  {journal} {\bibinfo
  {journal} {Computer Physics Communications}\ }\textbf {\bibinfo {volume}
  {207}},\ \bibinfo {pages} {35} (\bibinfo {year} {2016})}\BibitemShut
  {NoStop}%
\bibitem [{\citenamefont {Peeters}\ \emph {et~al.}(2009)\citenamefont
  {Peeters}, \citenamefont {Camenen}, \citenamefont {Casson}, \citenamefont
  {Hornsby}, \citenamefont {Snodin}, \citenamefont {Strintzi},\ and\
  \citenamefont {Szepesi}}]{Peeters_2009}%
  \BibitemOpen
  \bibfield  {author} {\bibinfo {author} {\bibfnamefont {A.}~\bibnamefont
  {Peeters}}, \bibinfo {author} {\bibfnamefont {Y.}~\bibnamefont {Camenen}},
  \bibinfo {author} {\bibfnamefont {F.}~\bibnamefont {Casson}}, \bibinfo
  {author} {\bibfnamefont {W.}~\bibnamefont {Hornsby}}, \bibinfo {author}
  {\bibfnamefont {A.}~\bibnamefont {Snodin}}, \bibinfo {author} {\bibfnamefont
  {D.}~\bibnamefont {Strintzi}},\ and\ \bibinfo {author} {\bibfnamefont
  {G.}~\bibnamefont {Szepesi}},\ }\bibfield  {title} {\bibinfo {title} {The
  nonlinear gyro-kinetic flux tube code {GKW}},\ }\href
  {https://doi.org/10.1016/j.cpc.2009.07.001} {\bibfield  {journal} {\bibinfo
  {journal} {Computer Physics Communications}\ }\textbf {\bibinfo {volume}
  {180}},\ \bibinfo {pages} {2650} (\bibinfo {year} {2009})}\BibitemShut
  {NoStop}%
\bibitem [{\citenamefont {Bourdelle}\ \emph {et~al.}(2007)\citenamefont
  {Bourdelle}, \citenamefont {Garbet}, \citenamefont {Imbeaux}, \citenamefont
  {Casati}, \citenamefont {Dubuit}, \citenamefont {Guirlet},\ and\
  \citenamefont {Parisot}}]{Bourdelle_2007}%
  \BibitemOpen
  \bibfield  {author} {\bibinfo {author} {\bibfnamefont {C.}~\bibnamefont
  {Bourdelle}}, \bibinfo {author} {\bibfnamefont {X.}~\bibnamefont {Garbet}},
  \bibinfo {author} {\bibfnamefont {F.}~\bibnamefont {Imbeaux}}, \bibinfo
  {author} {\bibfnamefont {A.}~\bibnamefont {Casati}}, \bibinfo {author}
  {\bibfnamefont {N.}~\bibnamefont {Dubuit}}, \bibinfo {author} {\bibfnamefont
  {R.}~\bibnamefont {Guirlet}},\ and\ \bibinfo {author} {\bibfnamefont
  {T.}~\bibnamefont {Parisot}},\ }\bibfield  {title} {\bibinfo {title} {A new
  gyrokinetic quasilinear transport model applied to particle transport in
  tokamak plasmas},\ }\href {https://doi.org/10.1063/1.2800869} {\bibfield
  {journal} {\bibinfo  {journal} {Physics of Plasmas}\ }\textbf {\bibinfo
  {volume} {14}},\ \bibinfo {pages} {112501} (\bibinfo {year}
  {2007})}\BibitemShut {NoStop}%
\bibitem [{\citenamefont {Citrin}\ \emph {et~al.}(2017)\citenamefont {Citrin},
  \citenamefont {Bourdelle}, \citenamefont {Casson}, \citenamefont {Angioni},
  \citenamefont {Bonanomi}, \citenamefont {Camenen}, \citenamefont {Garbet},
  \citenamefont {Garzotti}, \citenamefont {Görler}, \citenamefont {Gürcan},
  \citenamefont {Koechl}, \citenamefont {Imbeaux}, \citenamefont {Linder},
  \citenamefont {van~de Plassche}, \citenamefont {Strand},\ and\ \citenamefont
  {and}}]{Citrin_2017}%
  \BibitemOpen
  \bibfield  {author} {\bibinfo {author} {\bibfnamefont {J.}~\bibnamefont
  {Citrin}}, \bibinfo {author} {\bibfnamefont {C.}~\bibnamefont {Bourdelle}},
  \bibinfo {author} {\bibfnamefont {F.~J.}\ \bibnamefont {Casson}}, \bibinfo
  {author} {\bibfnamefont {C.}~\bibnamefont {Angioni}}, \bibinfo {author}
  {\bibfnamefont {N.}~\bibnamefont {Bonanomi}}, \bibinfo {author}
  {\bibfnamefont {Y.}~\bibnamefont {Camenen}}, \bibinfo {author} {\bibfnamefont
  {X.}~\bibnamefont {Garbet}}, \bibinfo {author} {\bibfnamefont
  {L.}~\bibnamefont {Garzotti}}, \bibinfo {author} {\bibfnamefont
  {T.}~\bibnamefont {Görler}}, \bibinfo {author} {\bibfnamefont
  {O.}~\bibnamefont {Gürcan}}, \bibinfo {author} {\bibfnamefont
  {F.}~\bibnamefont {Koechl}}, \bibinfo {author} {\bibfnamefont
  {F.}~\bibnamefont {Imbeaux}}, \bibinfo {author} {\bibfnamefont
  {O.}~\bibnamefont {Linder}}, \bibinfo {author} {\bibfnamefont
  {K.}~\bibnamefont {van~de Plassche}}, \bibinfo {author} {\bibfnamefont
  {P.}~\bibnamefont {Strand}},\ and\ \bibinfo {author} {\bibfnamefont {G.~S.}\
  \bibnamefont {and}},\ }\bibfield  {title} {\bibinfo {title} {Tractable
  flux-driven temperature, density, and rotation profile evolution with the
  quasilinear gyrokinetic transport model {QuaLiKiz}},\ }\href
  {https://doi.org/10.1088/1361-6587/aa8aeb} {\bibfield  {journal} {\bibinfo
  {journal} {Plasma Physics and Controlled Fusion}\ }\textbf {\bibinfo {volume}
  {59}},\ \bibinfo {pages} {124005} (\bibinfo {year} {2017})}\BibitemShut
  {NoStop}%
\bibitem [{\citenamefont {M.~Romanelli}(2014)}]{Romanelli_14}%
  \BibitemOpen
  \bibfield  {author} {\bibinfo {author} {\bibfnamefont {e.~a.}\ \bibnamefont
  {M.~Romanelli}},\ }\bibfield  {title} {\bibinfo {title} {Jintrac: A system of
  codes for integrated simulation of tokamak scenarios},\ }\href
  {https://doi.org/10.1585/pfr.9.3403023} {\bibfield  {journal} {\bibinfo
  {journal} {Plasma and Fusion Research}\ }\textbf {\bibinfo {volume} {9}},\
  \bibinfo {pages} {3403023} (\bibinfo {year} {2014})}\BibitemShut {NoStop}%
\bibitem [{\citenamefont {Hahm}\ \emph {et~al.}(2005)\citenamefont {Hahm},
  \citenamefont {Diamond}, \citenamefont {Lin}, \citenamefont {Rewoldt},
  \citenamefont {G{\"u}rcan},\ and\ \citenamefont {Ethier}}]{Hahm_PoP05}%
  \BibitemOpen
  \bibfield  {author} {\bibinfo {author} {\bibfnamefont {T.~S.}\ \bibnamefont
  {Hahm}}, \bibinfo {author} {\bibfnamefont {P.~H.}\ \bibnamefont {Diamond}},
  \bibinfo {author} {\bibfnamefont {Z.}~\bibnamefont {Lin}}, \bibinfo {author}
  {\bibfnamefont {G.}~\bibnamefont {Rewoldt}}, \bibinfo {author} {\bibfnamefont
  {{\"O}.~D.}\ \bibnamefont {G{\"u}rcan}},\ and\ \bibinfo {author}
  {\bibfnamefont {S.}~\bibnamefont {Ethier}},\ }\bibfield  {title} {\bibinfo
  {title} {On the dynamics of edge-core coupling},\ }\href
  {https://doi.org/10.1063/1.2034307} {\bibfield  {journal} {\bibinfo
  {journal} {Physics of Plasmas}\ }\textbf {\bibinfo {volume} {12}},\ \bibinfo
  {eid} {090903} (\bibinfo {year} {2005})}\BibitemShut {NoStop}%
\bibitem [{\citenamefont {Dif-Pradalier}\ \emph {et~al.}(2022)\citenamefont
  {Dif-Pradalier}, \citenamefont {Ghendrih}, \citenamefont {Sarazin},
  \citenamefont {Caschera}, \citenamefont {Clairet}, \citenamefont {Camenen},
  \citenamefont {Garbet}, \citenamefont {Grandgirard}, \citenamefont {Munschy},
  \citenamefont {Vermare},\ and\ \citenamefont
  {Widmer}}]{DifPradalier_encours22}%
  \BibitemOpen
  \bibfield  {author} {\bibinfo {author} {\bibfnamefont {G.}~\bibnamefont
  {Dif-Pradalier}}, \bibinfo {author} {\bibfnamefont {P.}~\bibnamefont
  {Ghendrih}}, \bibinfo {author} {\bibfnamefont {Y.}~\bibnamefont {Sarazin}},
  \bibinfo {author} {\bibfnamefont {E.}~\bibnamefont {Caschera}}, \bibinfo
  {author} {\bibfnamefont {F.}~\bibnamefont {Clairet}}, \bibinfo {author}
  {\bibfnamefont {Y.}~\bibnamefont {Camenen}}, \bibinfo {author} {\bibfnamefont
  {X.}~\bibnamefont {Garbet}}, \bibinfo {author} {\bibfnamefont
  {V.}~\bibnamefont {Grandgirard}}, \bibinfo {author} {\bibfnamefont
  {Y.}~\bibnamefont {Munschy}}, \bibinfo {author} {\bibfnamefont
  {L.}~\bibnamefont {Vermare}},\ and\ \bibinfo {author} {\bibfnamefont
  {F.}~\bibnamefont {Widmer}},\ }\bibfield  {title} {\bibinfo {title}
  {Transport in fusion plasmas: Is the tail wagging the dog?},\ }\bibfield
  {journal} {\bibinfo  {journal} {under review}\ }\href
  {https://doi.org/10.21203/rs.3.rs-879691/v1} {10.21203/rs.3.rs-879691/v1}
  (\bibinfo {year} {2022})\BibitemShut {NoStop}%
\bibitem [{\citenamefont {Lin}\ \emph {et~al.}(2002)\citenamefont {Lin},
  \citenamefont {Ethier}, \citenamefont {Hahm},\ and\ \citenamefont
  {Tang}}]{Lin_2002}%
  \BibitemOpen
  \bibfield  {author} {\bibinfo {author} {\bibfnamefont {Z.}~\bibnamefont
  {Lin}}, \bibinfo {author} {\bibfnamefont {S.}~\bibnamefont {Ethier}},
  \bibinfo {author} {\bibfnamefont {T.~S.}\ \bibnamefont {Hahm}},\ and\
  \bibinfo {author} {\bibfnamefont {W.~M.}\ \bibnamefont {Tang}},\ }\bibfield
  {title} {\bibinfo {title} {Size scaling of turbulent transport in
  magnetically confined plasmas},\ }\bibfield  {journal} {\bibinfo  {journal}
  {Physical Review Letters}\ }\textbf {\bibinfo {volume} {88}},\ \href
  {https://doi.org/10.1103/PhysRevLett.88.195004}
  {10.1103/PhysRevLett.88.195004} (\bibinfo {year} {2002})\BibitemShut
  {NoStop}%
\bibitem [{\citenamefont {Dimits}\ \emph {et~al.}(2000)\citenamefont {Dimits},
  \citenamefont {Bateman}, \citenamefont {Beer}, \citenamefont {Cohen},
  \citenamefont {Dorland}, \citenamefont {Hammett}, \citenamefont {Kim},
  \citenamefont {Kinsey}, \citenamefont {Kotschenreuther}, \citenamefont
  {Kritz}, \citenamefont {Lao}, \citenamefont {Mandrekas}, \citenamefont
  {Nevins}, \citenamefont {Parker}, \citenamefont {Redd}, \citenamefont
  {Shumaker}, \citenamefont {Sydora},\ and\ \citenamefont
  {Weiland}}]{Dimits_2000}%
  \BibitemOpen
  \bibfield  {author} {\bibinfo {author} {\bibfnamefont {A.~M.}\ \bibnamefont
  {Dimits}}, \bibinfo {author} {\bibfnamefont {G.}~\bibnamefont {Bateman}},
  \bibinfo {author} {\bibfnamefont {M.~A.}\ \bibnamefont {Beer}}, \bibinfo
  {author} {\bibfnamefont {B.~I.}\ \bibnamefont {Cohen}}, \bibinfo {author}
  {\bibfnamefont {W.}~\bibnamefont {Dorland}}, \bibinfo {author} {\bibfnamefont
  {G.~W.}\ \bibnamefont {Hammett}}, \bibinfo {author} {\bibfnamefont
  {C.}~\bibnamefont {Kim}}, \bibinfo {author} {\bibfnamefont {J.~E.}\
  \bibnamefont {Kinsey}}, \bibinfo {author} {\bibfnamefont {M.}~\bibnamefont
  {Kotschenreuther}}, \bibinfo {author} {\bibfnamefont {A.~H.}\ \bibnamefont
  {Kritz}}, \bibinfo {author} {\bibfnamefont {L.~L.}\ \bibnamefont {Lao}},
  \bibinfo {author} {\bibfnamefont {J.}~\bibnamefont {Mandrekas}}, \bibinfo
  {author} {\bibfnamefont {W.~M.}\ \bibnamefont {Nevins}}, \bibinfo {author}
  {\bibfnamefont {S.~E.}\ \bibnamefont {Parker}}, \bibinfo {author}
  {\bibfnamefont {A.~J.}\ \bibnamefont {Redd}}, \bibinfo {author}
  {\bibfnamefont {D.~E.}\ \bibnamefont {Shumaker}}, \bibinfo {author}
  {\bibfnamefont {R.}~\bibnamefont {Sydora}},\ and\ \bibinfo {author}
  {\bibfnamefont {J.}~\bibnamefont {Weiland}},\ }\bibfield  {title} {\bibinfo
  {title} {Comparisons and physics basis of tokamak transport models and
  turbulence simulations},\ }\href {https://doi.org/10.1063/1.873896}
  {\bibfield  {journal} {\bibinfo  {journal} {Physics of Plasmas}\ }\textbf
  {\bibinfo {volume} {7}},\ \bibinfo {pages} {969} (\bibinfo {year}
  {2000})}\BibitemShut {NoStop}%
\bibitem [{\citenamefont {Diamond}\ \emph {et~al.}(2010)\citenamefont
  {Diamond}, \citenamefont {Itoh},\ and\ \citenamefont {Itoh}}]{Diamond_2010}%
  \BibitemOpen
  \bibfield  {author} {\bibinfo {author} {\bibfnamefont {P.~H.}\ \bibnamefont
  {Diamond}}, \bibinfo {author} {\bibfnamefont {S.-I.}\ \bibnamefont {Itoh}},\
  and\ \bibinfo {author} {\bibfnamefont {K.}~\bibnamefont {Itoh}},\ }\href
  {https://www.ebook.de/de/product/10047272/patrick_h_diamond_modern_plasma_physics_volume_1_physical_kinetics_of_turbulent_plasmas.html}
  {\emph {\bibinfo {title} {Modern Plasma Physics: Volume 1, Physical Kinetics
  of Turbulent Plasmas}}}\ (\bibinfo  {publisher} {Cambridge University
  Press},\ \bibinfo {year} {2010})\BibitemShut {NoStop}%
\bibitem [{\citenamefont {Besse}\ \emph {et~al.}(2011)\citenamefont {Besse},
  \citenamefont {Elskens}, \citenamefont {Escande},\ and\ \citenamefont
  {Bertrand}}]{Besse_2011}%
  \BibitemOpen
  \bibfield  {author} {\bibinfo {author} {\bibfnamefont {N.}~\bibnamefont
  {Besse}}, \bibinfo {author} {\bibfnamefont {Y.}~\bibnamefont {Elskens}},
  \bibinfo {author} {\bibfnamefont {D.~F.}\ \bibnamefont {Escande}},\ and\
  \bibinfo {author} {\bibfnamefont {P.}~\bibnamefont {Bertrand}},\ }\bibfield
  {title} {\bibinfo {title} {Validity of quasilinear theory: refutations and
  new numerical confirmation},\ }\href
  {https://doi.org/10.1088/0741-3335/53/2/025012} {\bibfield  {journal}
  {\bibinfo  {journal} {Plasma Physics and Controlled Fusion}\ }\textbf
  {\bibinfo {volume} {53}},\ \bibinfo {pages} {025012} (\bibinfo {year}
  {2011})}\BibitemShut {NoStop}%
\bibitem [{\citenamefont {Casati}\ \emph {et~al.}(2009)\citenamefont {Casati},
  \citenamefont {Bourdelle}, \citenamefont {Garbet}, \citenamefont {Imbeaux},
  \citenamefont {Candy}, \citenamefont {Clairet}, \citenamefont
  {Dif-Pradalier}, \citenamefont {Falchetto}, \citenamefont {Gerbaud},
  \citenamefont {Grandgirard}, \citenamefont {Özgur. D.~Gürcan},
  \citenamefont {Hennequin}, \citenamefont {Kinsey}, \citenamefont {Ottaviani},
  \citenamefont {Sabot}, \citenamefont {Sarazin}, \citenamefont {Vermare},\
  and\ \citenamefont {Waltz}}]{Casati_2009}%
  \BibitemOpen
  \bibfield  {author} {\bibinfo {author} {\bibfnamefont {A.}~\bibnamefont
  {Casati}}, \bibinfo {author} {\bibfnamefont {C.}~\bibnamefont {Bourdelle}},
  \bibinfo {author} {\bibfnamefont {X.}~\bibnamefont {Garbet}}, \bibinfo
  {author} {\bibfnamefont {F.}~\bibnamefont {Imbeaux}}, \bibinfo {author}
  {\bibfnamefont {J.}~\bibnamefont {Candy}}, \bibinfo {author} {\bibfnamefont
  {F.}~\bibnamefont {Clairet}}, \bibinfo {author} {\bibfnamefont
  {G.}~\bibnamefont {Dif-Pradalier}}, \bibinfo {author} {\bibfnamefont
  {G.}~\bibnamefont {Falchetto}}, \bibinfo {author} {\bibfnamefont
  {T.}~\bibnamefont {Gerbaud}}, \bibinfo {author} {\bibfnamefont
  {V.}~\bibnamefont {Grandgirard}}, \bibinfo {author} {\bibnamefont {Özgur.
  D.~Gürcan}}, \bibinfo {author} {\bibfnamefont {P.}~\bibnamefont
  {Hennequin}}, \bibinfo {author} {\bibfnamefont {J.}~\bibnamefont {Kinsey}},
  \bibinfo {author} {\bibfnamefont {M.}~\bibnamefont {Ottaviani}}, \bibinfo
  {author} {\bibfnamefont {R.}~\bibnamefont {Sabot}}, \bibinfo {author}
  {\bibfnamefont {Y.}~\bibnamefont {Sarazin}}, \bibinfo {author} {\bibfnamefont
  {L.}~\bibnamefont {Vermare}},\ and\ \bibinfo {author} {\bibfnamefont {R.~E.}\
  \bibnamefont {Waltz}},\ }\bibfield  {title} {\bibinfo {title} {Validating a
  quasi-linear transport model versus nonlinear simulations},\ }\href
  {https://doi.org/10.1088/0029-5515/49/8/085012} {\bibfield  {journal}
  {\bibinfo  {journal} {Nuclear Fusion}\ }\textbf {\bibinfo {volume} {49}},\
  \bibinfo {pages} {085012} (\bibinfo {year} {2009})}\BibitemShut {NoStop}%
\bibitem [{\citenamefont {Heinonen}\ and\ \citenamefont
  {Diamond}(2019)}]{Heinonen_PoP19}%
  \BibitemOpen
  \bibfield  {author} {\bibinfo {author} {\bibfnamefont {R.~A.}\ \bibnamefont
  {Heinonen}}\ and\ \bibinfo {author} {\bibfnamefont {P.~H.}\ \bibnamefont
  {Diamond}},\ }\bibfield  {title} {\bibinfo {title} {Subcritical turbulence
  spreading and avalanche birth},\ }\href {https://doi.org/10.1063/1.5083176}
  {\bibfield  {journal} {\bibinfo  {journal} {Physics of Plasmas}\ }\textbf
  {\bibinfo {volume} {26}},\ \bibinfo {pages} {030701} (\bibinfo {year}
  {2019})},\ \Eprint {https://arxiv.org/abs/https://doi.org/10.1063/1.5083176}
  {https://doi.org/10.1063/1.5083176} \BibitemShut {NoStop}%
\bibitem [{\citenamefont {Baschetti}\ \emph {et~al.}(2021)\citenamefont
  {Baschetti}, \citenamefont {Bufferand}, \citenamefont {Ciraolo},
  \citenamefont {Ghendrih}, \citenamefont {Serre}, \citenamefont {Tamain},\
  and\ \citenamefont {the WEST~Team}}]{Baschetti_NF21}%
  \BibitemOpen
  \bibfield  {author} {\bibinfo {author} {\bibfnamefont {S.}~\bibnamefont
  {Baschetti}}, \bibinfo {author} {\bibfnamefont {H.}~\bibnamefont
  {Bufferand}}, \bibinfo {author} {\bibfnamefont {G.}~\bibnamefont {Ciraolo}},
  \bibinfo {author} {\bibfnamefont {P.}~\bibnamefont {Ghendrih}}, \bibinfo
  {author} {\bibfnamefont {E.}~\bibnamefont {Serre}}, \bibinfo {author}
  {\bibfnamefont {P.}~\bibnamefont {Tamain}},\ and\ \bibinfo {author}
  {\bibnamefont {the WEST~Team}},\ }\bibfield  {title} {\bibinfo {title}
  {Self-consistent cross-field transport model for core and edge plasma
  transport},\ }\href {https://doi.org/10.1088/1741-4326/ac1e60} {\bibfield
  {journal} {\bibinfo  {journal} {Nuclear Fusion}\ }\textbf {\bibinfo {volume}
  {61}},\ \bibinfo {pages} {106020} (\bibinfo {year} {2021})}\BibitemShut
  {NoStop}%
\bibitem [{\citenamefont {Garbet}\ \emph {et~al.}(2007)\citenamefont {Garbet},
  \citenamefont {Sarazin}, \citenamefont {Imbeaux}, \citenamefont {Ghendrih},
  \citenamefont {Bourdelle}, \citenamefont {Gurcan},\ and\ \citenamefont
  {Diamond}}]{Garbet_PoP07}%
  \BibitemOpen
  \bibfield  {author} {\bibinfo {author} {\bibfnamefont {X.}~\bibnamefont
  {Garbet}}, \bibinfo {author} {\bibfnamefont {Y.}~\bibnamefont {Sarazin}},
  \bibinfo {author} {\bibfnamefont {F.}~\bibnamefont {Imbeaux}}, \bibinfo
  {author} {\bibfnamefont {P.}~\bibnamefont {Ghendrih}}, \bibinfo {author}
  {\bibfnamefont {C.}~\bibnamefont {Bourdelle}}, \bibinfo {author}
  {\bibfnamefont {O.~D.}\ \bibnamefont {Gurcan}},\ and\ \bibinfo {author}
  {\bibfnamefont {P.~H.}\ \bibnamefont {Diamond}},\ }\bibfield  {title}
  {\bibinfo {title} {Front propagation and critical gradient transport
  models},\ }\href {https://doi.org/10.1063/1.2824375} {\bibfield  {journal}
  {\bibinfo  {journal} {Physics of Plasmas}\ }\textbf {\bibinfo {volume}
  {14}},\ \bibinfo {pages} {122305} (\bibinfo {year} {2007})},\ \Eprint
  {https://arxiv.org/abs/https://doi.org/10.1063/1.2824375}
  {https://doi.org/10.1063/1.2824375} \BibitemShut {NoStop}%
\bibitem [{\citenamefont {Dominski}\ \emph {et~al.}(2012)\citenamefont
  {Dominski}, \citenamefont {Brunner}, \citenamefont {Aghdam}, \citenamefont
  {G\"orler}, \citenamefont {Jenko},\ and\ \citenamefont
  {Told}}]{Dominski_2012}%
  \BibitemOpen
  \bibfield  {author} {\bibinfo {author} {\bibfnamefont {J.}~\bibnamefont
  {Dominski}}, \bibinfo {author} {\bibfnamefont {S.}~\bibnamefont {Brunner}},
  \bibinfo {author} {\bibfnamefont {S.~K.}\ \bibnamefont {Aghdam}}, \bibinfo
  {author} {\bibfnamefont {T.}~\bibnamefont {G\"orler}}, \bibinfo {author}
  {\bibfnamefont {F.}~\bibnamefont {Jenko}},\ and\ \bibinfo {author}
  {\bibfnamefont {D.}~\bibnamefont {Told}},\ }\bibfield  {title} {\bibinfo
  {title} {Identifying the role of non-adiabatic passing electrons in itg/tem
  microturbulence by comparing fully kinetic and hybrid electron simulations},\
  }\href {http://stacks.iop.org/1742-6596/401/i=1/a=012006} {\bibfield
  {journal} {\bibinfo  {journal} {Journal of Physics: Conference Series}\
  }\textbf {\bibinfo {volume} {401}},\ \bibinfo {pages} {012006} (\bibinfo
  {year} {2012})}\BibitemShut {NoStop}%
\bibitem [{\citenamefont {Rath}\ \emph {et~al.}(2021)\citenamefont {Rath},
  \citenamefont {Peeters},\ and\ \citenamefont {Weikl}}]{Rath_PoP21}%
  \BibitemOpen
  \bibfield  {author} {\bibinfo {author} {\bibfnamefont {F.}~\bibnamefont
  {Rath}}, \bibinfo {author} {\bibfnamefont {A.~G.}\ \bibnamefont {Peeters}},\
  and\ \bibinfo {author} {\bibfnamefont {A.}~\bibnamefont {Weikl}},\ }\bibfield
   {title} {\bibinfo {title} {Analysis of zonal flow pattern formation and the
  modification of staircase states by electron dynamics in gyrokinetic near
  marginal turbulence},\ }\href {https://doi.org/10.1063/5.0054358} {\bibfield
  {journal} {\bibinfo  {journal} {Physics of Plasmas}\ }\textbf {\bibinfo
  {volume} {28}},\ \bibinfo {pages} {072305} (\bibinfo {year} {2021})},\
  \Eprint {https://arxiv.org/abs/https://doi.org/10.1063/5.0054358}
  {https://doi.org/10.1063/5.0054358} \BibitemShut {NoStop}%
\end{thebibliography}

%

\end{document}